\documentclass[11pt]{article} 
\usepackage[utf8]{inputenc}
\usepackage{multirow}
\usepackage{amsfonts}
\usepackage{amsmath}
\usepackage{amssymb}
\usepackage{amsthm}
\usepackage[dvipsnames]{xcolor}
    \definecolor{darkblue}{rgb}{0,0.5,0}
    \definecolor{darkblue}{rgb}{0,0,0.6}
    \definecolor{purple}{rgb}{0.4,.2,0.7}
\usepackage[margin = 2.5cm]{geometry}
\usepackage{graphicx}
\usepackage[hyperfootnotes = false, colorlinks = true, linkcolor = darkblue, citecolor = purple]{hyperref}
\usepackage{subcaption}
\usepackage{ytableau}

\usepackage{csquotes}

\usepackage{tikz}
\usetikzlibrary{decorations.pathmorphing}
\usetikzlibrary{decorations.markings}
\definecolor{mathblue}{RGB}{180,44,37}
\definecolor{mathblue}{RGB}{39,94,190}
\tikzset{>=latex} 
\tikzset{ photon/.style={decorate, decoration={snake}, draw=black}}

\usepackage{comment}

\usepackage{physics}

\newcommand{\be}{\begin{equation}}
\newcommand{\ee}{\end{equation}}
\newcommand{\bea}{\begin{eqnarray}}
\newcommand{\eea}{\end{eqnarray}}

\begin{document}

\thispagestyle{empty}
\begin{center}
    ~\vspace{5mm}

  \vskip 2cm 
  
   {\LARGE \bf 
       Feynman-like parameterizations of (anti-)de Sitter Witten diagrams for all masses at any loop order
   }

   \vspace{0.5in}
     
   {\bf Aidan Herderschee
   }

    \vspace{0.5in}

 Institute for Advanced Study, Princeton, NJ 08540, USA
                
    \vspace{0.5in}

    \vspace{0.5in}
    

\end{center}

\vspace{0.5in}

\begin{abstract}
Computing correlation functions in curved spacetime is central to both theoretical and experimental efforts, from precision cosmology to quantum simulations of strongly coupled systems. In anti-de Sitter (AdS) and de Sitter (dS) space, the key observables, boundary correlators in AdS and late-time correlators in dS, are obtained via Witten diagram calculations. While formally analogous to flat-space Feynman diagrams, even tree-level Witten diagrams are significantly more complicated due to the structure of bulk propagators. Existing computational approaches often focus on scalars with a specific “conformal” mass, for which propagators simplify enough to permit the use of standard flat-space techniques. This restriction, however, omits the generic internal-line masses that arise in many cosmological and holographic settings. We present the Witten-Feynman (WF) parameterization, a general representation of scalar Witten diagrams in (A)dS as generalized Euler integrals. The WF framework applies in both position and momentum space, accommodates arbitrary internal and external masses, and holds at any loop order. It directly generalizes the familiar Feynman parameterization form of Feynman integrals, making it possible to import a broad range of amplitude techniques into the curved-space setting. We illustrate the method through two applications: a generalization of Weinberg’s theorem on ultraviolet convergence and a series expansion technique that can yield explicit evaluations. Our results provide a unified computational tool for (A)dS boundary correlators, opening the door to more systematic calculations relevant for upcoming experiments and simulations.
\end{abstract}

\vspace{1in}

\pagebreak

\setcounter{tocdepth}{3}
{\hypersetup{linkcolor=black}\tableofcontents}

\section{Introduction}\label{sec:introduction}

The universe is not flat. The deviation from flatness is slight, but the universe’s geometry is
curved, bending under the weight of its contents. Observations of the cosmic microwave background (CMB), gravitational lensing, and the large-scale distribution of galaxies provide direct evidence for curvature. Photons from the early universe propagate along null geodesics in a curved background, imprinting subtle but detectable distortions in the CMB temperature map \cite{Planck:2018lbu}, and the redshift of receding galaxies encode the underlying geometry \cite{gardner2023james,rigby2023science}.\footnote{Within the $\Lambda$CDM model, these measurements yield differing estimates of the Hubble constant \cite{DiValentino:2021izs}.} Such effects are not confined to cosmology: even the operation of certain GPS satellite systems must account for general-relativistic curvature. Over the course of a single day, neglecting the gravitational redshift from Earth’s field alone would cause timing errors that translate into navigational inaccuracies of about $10$ kilometers in GPS systems that rely on precise synchronization between Earth-based and satellite clocks \cite{ashby1995relativistic}. To reproduce reality with any fidelity, our theories must treat curvature not as a minor perturbation, but a fundamental facet of nature.

Unfortunately, computing observables in a generic curved spacetime is hard. We therefore focus on the two maximally symmetric cases, anti-de Sitter (AdS) and de Sitter (dS), which are nonetheless of significant experimental relevance. In the context of AdS/CFT \cite{Maldacena:1997re,Witten:1998qj,Gubser:1998bc}, gravitational theories in AdS are conjectured to be dual to specific strongly coupled quantum field theories; boundary correlators in the gravitational description then yield concrete predictions for the dual non-gravitational systems, testable via Monte Carlo calculations \cite{Catterall:2007fp,Hanada:2007ti,Anagnostopoulos:2007fw,Catterall:2008yz,Hanada:2008ez,Hanada:2008gy,Catterall:2009xn,Hanada:2011fq,Filev:2015hia,Berkowitz:2016jlq,Berkowitz:2018qhn,Bergner:2021goh,Pateloudis:2022ijr} and emerging quantum-simulation platforms \cite{Rinaldi:2021jbg,Maldacena:2023acv,Halimeh:2024bth}. In dS, late-time (``boundary'') correlators are likewise central: inflation predicts a pre-reheating epoch well approximated by dS \cite{Guth:1980zm,Linde:1981mu,Albrecht:1982wi,Starobinsky:1982ee}, leading to sharp observational signatures \cite{Maldacena:2002vr,Achucarro:2022qrl}. Primordial four-point correlation functions (the trispectrum) can, in some cases, compete with three-point correlation functions (the bispectrum) due to enhanced signal-to-noise scaling, suggesting that higher point correlators may be relevant for future measurements \cite{Kalaja:2020mkq}.\footnote{See Refs. \cite{Kogo:2006kh,Creminelli:2006gc,Bordin:2019tyb} for some relevant papers.} Thus explicit expressions for boundary correlators in (A)dS are relevant for a diverse set of future experimental programs, from quantum simulation platforms to precision cosmological measurements.

Realizing this opportunity requires efficient methods to compute boundary correlators. We draw on the substantial progress in scattering-amplitude technology \cite{Travaglini:2022uwo}, the flat-space analog of boundary correlators. Amplitudes admit algorithmic evaluation via Feynman diagrams and remain tractable at low orders in perturbation theory. For example, the $s$-channel diagram in $\phi^3$ theory yields a simple rational function:
\begin{equation}
\pgfmathsetmacro{\r}{1}
\begin{tikzpicture}[baseline={([yshift=-.5ex]current bounding box.center)},every node/.style={font=\scriptsize}]
\filldraw (45:\r) circle (1pt) node[above=0pt]{$3$};
\filldraw (135:\r) circle (1pt) node[above=0pt]{$2$};
\filldraw (225:\r) circle (1pt) node[below=0pt]{$1$};
\filldraw (-45:\r) circle (1pt) node[below=0pt]{$4$};
\filldraw (\r/2,0) circle (1pt) (0:\r/2) circle (1pt) (180:\r/2) circle (1pt);
\draw [thick] (45:\r) -- (0:\r/2) -- (-45:\r);
\draw [thick] (135:\r) -- (180:\r/2) -- (225:\r);
\draw [thick] (0:\r/2) -- (180:\r/2);
\end{tikzpicture}
\propto \frac{1}{(p_1 + p_2)^2}\,.
\end{equation}
One-loop Feynman integrals can often be expressed in polylogarithms \cite{Henn:2013pwa,Bourjaily:2019exo,Herrmann:2019upk,Chen:2020uyk}, with serious computational challenges typically arising at two loops and beyond. In (A)dS, by contrast, boundary correlators are computed via Witten diagrams \cite{Witten:1998qj,Freedman:1998tz,DHoker:1999mqo}, and even tree-level results are substantially more intricate. The core complication is the bulk two-point propagator $G_{\Delta}(X_{\mathrm{AdS}};X_{\mathrm{AdS}}')$, which involves a ${}_2F_1$ function with $\Delta$ set by the mass. For instance, the four-point $s$-channel diagram involves a double integral over this propagator:
\begin{align}\label{4pointscala}
\pgfmathsetmacro{\r}{1}
\begin{tikzpicture}[baseline={([yshift=-.5ex]current bounding box.center)},every node/.style={font=\scriptsize}]
\draw [] (0,0) circle (\r cm);
\filldraw (45:\r) circle (1pt) node[above=0pt]{$3$};
\filldraw (135:\r) circle (1pt) node[above=0pt]{$2$};
\filldraw (225:\r) circle (1pt) node[below=0pt]{$1$};
\filldraw (-45:\r) circle (1pt) node[below=0pt]{$4$};
\filldraw (\r/2,0) circle (1pt) (0:\r/2) circle (1pt) (180:\r/2) circle (1pt);
\draw [thick] (45:\r) -- (0:\r/2) -- (-45:\r);
\draw [thick] (135:\r) -- (180:\r/2) -- (225:\r);
\draw [thick] (0:\r/2) -- (180:\r/2);
\end{tikzpicture}
\propto &\int dX_{\textrm{AdS}}\sqrt{g_{\textrm{AdS}}(X_{\textrm{AdS}})}dX_{\textrm{AdS}}'\sqrt{g_{\textrm{AdS}}(X_{\textrm{AdS}}')} \, K_{\Delta_1}(X_{\textrm{AdS}}; Y_{1,\partial \textrm{AdS}})  \\
&\times K_{\Delta_2}(X_{\textrm{AdS}}; Y_{2,\partial \textrm{AdS}})G_{\Delta_I}(X_{\textrm{AdS}};X_{\textrm{AdS}}') \, K_{\Delta_3}(X_{\textrm{AdS}}'; Y_{3,\partial \textrm{AdS}})\, K_{\Delta_4}(X_{\textrm{AdS}}'; Y_{4,\partial \textrm{AdS}})\,, \nonumber
\end{align}
where the bulk-to-boundary propagators $K_{\Delta}$ are much simpler than $G_{\Delta}$. In special cases (particular dimensions and masses), Eq.~\eqref{4pointscala} reduces to polylogarithms \cite{DHoker:1999mqo}, but in general defines a more complicated function.

Our goal is to import amplitude methods into (A)dS correlator computations. To that end, we introduce Witten-Feynman (WF) parameterizations: new representations of (A)dS Witten diagrams as generalized Euler integrals \cite{Matsubara-Heo:2023ylc}.\footnote{A generalized Euler integral has the schematic form
\[
\int_{\mathcal{C}} d\alpha_{1}\cdots d\alpha_{n}\ \prod_{j=1}^{s} P_{j}(\alpha_{i},z_{i})^{\lambda_{j}},
\]
with polynomials $P_j$ and constants $\lambda_j$.} The WF framework applies in both position and momentum space, for scalar diagrams with arbitrary internal and external masses and at any loop order. It directly generalizes the familiar Feynman parameterization of Feynman integrals,
\begin{equation}\label{feynmanpara}
\text{(any Feynman integral)}\ \propto\  \int \!\Big[\prod_{e\in E^{\mathrm{int}}}\!d\alpha_{e}\Big]\,
\frac{\mathcal{U}(\alpha_{e})^{\#}}{\mathcal{F}(\alpha_{e},k_{i})^{\#}}\,
\delta\!\big(1-\sum_{e\in E}\alpha_{e}\big),
\end{equation}
where $\mathcal{U}$ and $\mathcal{F}$ are the Symanzik polynomials and $\#$ denotes constants fixed by masses and dimension; see \cite{Weinzierl:2022eaz}. The Feynman parameterization is a ubiquitous tool in flat-space quantum field theory, appearing in almost every introductory text due to its utility in organizing and evaluating loop integrals \cite{Peskin:1995ev,Weinberg:1995mt,Srednicki:2007qs,Schwartz:2014sze}. Its generalization to Witten diagrams in (A)dS space, therefore, holds promise as a similarly powerful framework, one that may streamline computations and reveal new structural insights in (A)dS. For simplicity, we restrict to Witten diagrams of non-derivatively coupled scalars in the majority of this paper, discussing the generalization to derivatively coupled particles and spinning propagators in section \ref{discussion}. 

Many works have explored the relationship between Feynman integrals and Witten diagrams \cite{Heckelbacher:2022fbx,Heckelbacher:2022hbq,Chowdhury:2023khl,Arkani-Hamed:2023kig,Arkani-Hamed:2024jbp,Chestnov:2025whi}, most often in the simplified setting of scalars with a special ``conformal'' mass, for which the two-point functions take a particularly tractable form.\footnote{Some exceptions include \cite{Caloro:2023cep,Liu:2024str,Alaverdian:2024llo,Biggs:2025qfh,Bobev:2025idz}.}\footnote{The conformal mass renders the free scalar theory conformally invariant, allowing the (A)dS two-point function to be obtained via a conformal map from flat space.} This restriction is partly pragmatic. Starting from Witten diagrams with external scalars with conformal masses, results for arbitrary \emph{external} masses can be obtained by applying differential operators to the external legs \cite{Arkani-Hamed:2018kmz,Benincasa:2019vqr}. However, such operators leave the \emph{internal} masses unchanged, and diagrams with non-conformal internal propagators, common in both cosmological and holographic contexts, require more general methods.

We now sketch the WF representations. In position space, Witten diagrams with generic scalar masses take the generalized Euler form
\begin{equation}\label{WFrepsketch}
\begin{split}
\pgfmathsetmacro{\r}{1}
\begin{tikzpicture}[baseline={([yshift=-.5ex]current bounding box.center)},every node/.style={font=\scriptsize}]
\draw [] (0,0) circle (\r cm);
\filldraw (0,0) circle (0pt) node{Anything};
\end{tikzpicture} \Bigg|_{\textrm{Position Space}} \propto & \int \prod_{e\in E^{\textrm{ext}}} d\alpha_{e} \,\alpha_{e}^{\#}\!
\prod_{e\in E^{\textrm{int}}} d\beta_{e}\,d\kappa_{e}\, \beta_{e}^{\#}\kappa_{e}^{\#}(\beta_{e}-\kappa_{e})^{\#}
\prod_{v\in V^{\textrm{int}}} dz_{v}\,z_{v}^{\#} \\
& \times \mathcal{W}^{\#}\,\mathcal{Y}^{\#}\,\mathcal{Z}^{\#}\,
\delta\!\Big(1-\sum \alpha_{e}-\sum (\beta_{e}+\kappa_{e})\Big)\,
\delta\!\Big(1-\sum z_{v}\Big),
\end{split}
\end{equation}
where $E^{\textrm{int}}$ and $E^{\textrm{ext}}$ are the sets of internal and external edges and $V^{\textrm{int}}$ the internal vertices. The polynomials $\mathcal{Y},\mathcal{W},\mathcal{Z}$ depend on the integration variables and on external positions; exponents are fixed by masses, dimension, and whether the space is AdS or dS (explicit formulas in Section~\ref{poswf}). As an example, the AdS four-point $s$-channel graph reads
\begin{equation}
\begin{split}
\pgfmathsetmacro{\r}{1}
\begin{tikzpicture}[baseline={([yshift=-.5ex]current bounding box.center)},every node/.style={font=\scriptsize}]
\draw [] (0,0) circle (\r cm);
\filldraw (45:\r) circle (1pt) node[above=0pt]{$3$};
\filldraw (135:\r) circle (1pt) node[above=0pt]{$2$};
\filldraw (225:\r) circle (1pt) node[below=0pt]{$1$};
\filldraw (-45:\r) circle (1pt) node[below=0pt]{$4$};
\filldraw (\r/2,0) circle (1pt) (0:\r/2) circle (1pt) (180:\r/2) circle (1pt);
\draw [thick] (45:\r) -- (0:\r/2) -- (-45:\r);
\draw [thick] (135:\r) -- (180:\r/2) -- (225:\r);
\draw [thick] (0:\r/2) -- (180:\r/2) ;
\end{tikzpicture}&\propto
\left[\int_{0}^{\infty}\prod_{i=1}^{4}\frac{d\alpha_{i}}{\alpha_{i}}\alpha_{i}^{\Delta_{i}}\right]
\int_{0}^{\infty} d\kappa \int _{\kappa}^{\infty}d\beta\ \beta^{d-\Delta_{I}-1}\,(\kappa(\beta-\kappa))^{\Delta_{I}-\frac{d+1}{2}} \\ 
&\times \int \frac{dz_{1}}{z_{1}}\frac{dz_{2}}{z_{2}}\, z_{1}^{\Delta_{1}+\Delta_{2}+\Delta_{I}-d}\,z_{2}^{\Delta_{3}+\Delta_{4}+\Delta_{I}-d}\,
\frac{\mathcal{Y}^{\frac{\Delta^{\textrm{ext}}-d}{2}}_{s}}{\mathcal{W}_{s}^{\frac{\Delta^{\textrm{ext}}}{2}}}\,
\mathcal{Z}_{s}^{\,d-\Delta_{I}-\frac{\Delta^{\textrm{ext}}}{2}} \\
&\times \delta\!\Big(1-\beta-\kappa-\sum_{i=1}^{4}\alpha_{i}\Big)\,\delta(1-z_{1}-z_{2}),
\end{split}
\end{equation}
with
\begin{equation}\label{schannelpoly}
\begin{split}
\Delta^{\textrm{ext}}&=\sum_{i=1}^{4}\Delta_{i}, \\
\mathcal{W}_{s}&=\alpha _1 \alpha _2 \beta _1 x_{1,2}^{2}+\alpha _1 \alpha _3 \beta _1 x_{1,3}^{2}+\alpha _1 \alpha _4 \beta _1 x_{1,4}^{2}+\alpha _1 \alpha _2 \alpha _3 x_{1,2}^{2}+\alpha _1 \alpha _2 \alpha _4 x_{1,2}^{2} \\
&\quad+\alpha _2 \alpha _3 \beta _1 x_{2,3}^{2}+\alpha _2 \alpha _4 \beta _1 x_{2,4}^{2}+\alpha _3 \alpha _4 \beta _1 x_{3,4}^{2}+\alpha _1 \alpha _3 \alpha _4 x_{3,4}^{2}+\alpha _2 \alpha _3 \alpha _4 x_{3,4}^{2}, \\
\mathcal{Y}_{s}&=\alpha _3 \beta _1+\alpha _1 \beta _1+\alpha _2 \beta _1+\alpha _4 \beta _1+\alpha _1 \alpha _3+\alpha _2 \alpha _3+\alpha _1 \alpha _4+\alpha _2 \alpha _4, \\
\mathcal{Z}_{s}&=\alpha_{1}z_{1}^{2}+\alpha_{2}z_{1}^{2}+\alpha_{3}z_{2}^{2}+\alpha_{4}z_{2}^{2}+\beta_{1}(z_{1}-z_{2})^{2}+4 \kappa_{1} z_{1}z_{2}.
\end{split}
\end{equation}
After using the two delta functions, there are six independent integration variables. Conformal symmetry reduces the six $x_{ij}$ to two cross-ratios, see Section~\ref{levconf}; in general an $n$-point conformal diagram depends on $n(n-3)/2$ cross-ratios.

A parallel representation holds in momentum space that more closely resembles flat-space:
\begin{equation}
\begin{split}
\pgfmathsetmacro{\r}{1}
\begin{tikzpicture}[baseline={([yshift=-.5ex]current bounding box.center)},every node/.style={font=\scriptsize}]
\draw [] (0,0) circle (\r cm);
\filldraw (0,0) circle (0pt) node{Anything};
\end{tikzpicture}\Bigg|_{\textrm{Momentum Space}} &\propto  \int \prod_{e\in E^{\textrm{ext}}} d\tilde{\alpha}_{e} \,\tilde{\alpha}_{e}^{\#}
\prod_{e\in E^{\textrm{int}}} d\tilde{\beta}_{e}\,d\tilde{\kappa}_{e}\, \tilde{\beta}_{e}^{\#}\tilde{\kappa}_{e}^{\#}(\tilde{\kappa}_{e}-\tilde{\beta}_{e})^{\#}
\prod_{v\in V} dz_{v}\,z_{v}^{\#} \\
& \times \mathcal{U}^{\#}(\mathcal{S}\mathcal{U}+\mathcal{F})^{\#}\,\tilde{\mathcal{Z}}^{\#}\,
\delta\!\Big(1-\sum \tilde{\alpha}_{e}-\sum (\tilde{\beta}_{e}+\tilde{\kappa}_{e})\Big)\,\delta\!\Big(1-\sum z_{v}\Big),
\end{split}
\end{equation}
where $\mathcal{U}$ and $\mathcal{F}$ are the usual Symanzik polynomials and $\mathcal{S}$ is new (explicit expressions in Section~\ref{wittenrevmom}). Passing between AdS and dS leaves the polynomials unchanged but modifies exponents and contours.

We demonstrate the utility of the WF parameterizations through two concrete applications. In Section \ref{wittenrevmomUV}, we show how the momentum-space WF parameterization allows us to prove a generalization of Weinberg's theorem on the ultraviolet (UV) convergence of Feynman integrals \cite{Weinberg:1959nj}. In Section~\ref{diffeqAdS}, we show how to convert the WF parameterized integrals into an infinite series that, at least in principle, offers a computationally inexpensive method for evaluating the integral explicitly. These results serve as a proof of concept, and we emphasize that many additional applications of the WF parameterization are possible; some of these are discussed in Section~\ref{discussion}. \\

\noindent Notation: We summarize some notation that we use throughout the text here for the reader's convenience. 
\begin{itemize}
\item $V^{\textrm{ext}}$ is the set of external vertices of the relevant diagram. External vertices are those that lie on the circle containing the Witten diagram. They correspond to boundary operator insertions. 
\item $V^{\textrm{int}}$ is the set of internal vertices of the relevant diagram. Internal vertices lie inside the circle enclosing the Witten diagram and correspond to bulk points that we integrate over.
\item $E^{\textrm{ext}}$ is the set of external edges of the relevant diagram. External edges are edges that connect to an external vertex. 
\item $E^{\textrm{int}}$ is the set of internal edges of the relevant diagram. Internal edges are edges that do not connect to an external vertex.
\item $E$ is the set of all edges, both internal and external. 
\item $\beta_{e}$ and $\kappa_{e}$ ($\tilde{\beta}_{e}$ and $\tilde{\kappa}_{e}$) are position (momentum) space Feynman parameters associated with the bulk-to-bulk propagator of edge $e$.
\item $\alpha_{e}$ ($\tilde{\alpha}_{e}$) are the position (momentum) space Feynman parameters associated with the bulk-to-bulk propagator of edge $e$.
\item $\Delta^{\textrm{ext}}$ is the sum of conformal dimensions, $\Delta_{i}$, of the external states.
\item $\Delta^{\textrm{int}}$ is the sum of conformal dimensions, $\Delta_{i}$, of the internal propagators.
\item $|\ldots|$ denotes the size of the set. 
\item $d+1$ is the spacetime dimension
\item $x_{i,j}^{2}=(x_{i}-x_{j})^{2}$
\end{itemize}
\vspace{\baselineskip}
\noindent \textit{Note added}: While this paper was in preparation, Ref. \cite{Raman:2025tsg} appeared, which also considered the evaluation of general Witten diagrams via series expansion.

\section{Review: Witten diagrams}\label{wittenrev}

In this section, we briefly review Witten diagrams in (A)dS. Following standard notation in the AdS/CFT literature, we use $d$ to denote the spacetime dimension of the boundary, not the bulk, which has spacetime dimension $d+1$. See Ref. \cite{Penedones:2016voo} for a review of AdS/CFT and Ref. \cite{Spradlin:2001pw} for a review of QFT in dS.

\subsection{Anti-de Sitter}\label{adsrev}

In AdS, we will work in the Pioncare patch with the metric: 
\begin{equation}
ds^{2}=\frac{dz^{2}+g_{\mu\nu}^{\textrm{M}}dx^{\mu}dx^{\nu}}{z^{2}} \ ,
\end{equation}
where $g_{\mu\nu}^{\textrm{M}}$ is the Minkowski metric on $\mathbb{R}^{d}$.\footnote{We note the above coordinates do not make the isometries of AdS manifest, unlike embedding space \cite{Penedones:2016voo}, but is nonetheless sufficient for our purpose.} Two-point propagators, which serve as the building blocks of Witten diagrams, satisfy the differential equation
\begin{equation}
\left [ z^{2}\partial_{z}^{2}-2z\partial_{z}+z^{2}g^{\mu\nu,\textrm{M}}\frac{\partial}{\partial x^{\mu}}\frac{\partial}{\partial x^{\nu}}-m^{2} \right ] G_{\Delta}(z,x^{\mu};z',x^{\mu}{}')=z^{d+1}\delta(z-z')\delta^{d}(x-x') \ , 
\end{equation}
which admits the solution 

\begin{equation}\label{adesitterprop}
G_{\Delta}^{\textrm{AdS}}(z,x^{\mu};z',x^{\mu}{}')=\frac{C_{\Delta}}{2(\Delta-\frac{d}{2})\zeta^{\Delta}}\ _{2}F_{1}\left(\Delta,\Delta-\frac{d}{2}+\frac{1}{2},2\Delta-d+1,-\frac{4}{\zeta}\right) \ ,
\end{equation}
where
\begin{equation}
\begin{split}
C_{\Delta}&=\frac{\Gamma(\Delta)}{\pi^{d/2}\Gamma(\Delta-d/2)}  \ , \\
\zeta&=\frac{(z-z')^{2}+(x-x')^{2}}{zz'} \ , \\
m^{2}&=\Delta(\Delta-d)\ .
\end{split}
\end{equation}
$\Delta$ corresponds to the conformal scaling dimension of the dual operator in the boundary theory. When the operator approaches the boundary, the propagator simplifies to
\begin{equation}\label{eq:adsboundprop}
\lim_{z'\rightarrow 0}G_{\Delta}^{\textrm{AdS}}(z,x^{\mu};z',x^{\mu}{}') =\frac{z'^{\Delta}}{2\Delta-d}K_{\Delta}^{\textrm{AdS}}(z,x;x'), \quad K_{\Delta}^{\textrm{AdS}}(z,x;x')=C_{\Delta}\left ( \frac{z}{z^{2}+(x-x')^{2}} \right)^{\Delta} \ .
\end{equation}
We will restrict to scalar theories for simplicity, but the techniques in this paper trivially generalize to theories with higher-spin particles.

We consider only operators inserted near the boundary, $z \rightarrow 0$, with the normalization:
\begin{equation}
\lim_{z\rightarrow 0}\mathcal{O}_{\textrm{bulk}}(z,x)=z^{-\Delta}(2\Delta-d)\mathcal{O}_{\textrm{boundary}}(x) \ .
\end{equation}
The perturbative computation of correlators in (A)dS involves summing over all relevant diagrams, Witten diagrams, from the set of Feynman rules and computing the corresponding integrals. Our goal is to calculate individual Witten diagrams such as Eq. (\ref{4pointscala}). Each edge in a Witten diagram maps to a propagator. Bulk vertices correspond to bulk points in AdS that we integrate over. For example, Eq.~(\ref{4pointscala}) gives the s-channel Witten diagram. An example of a five-point graph is 
\begin{align}\label{5point}
\pgfmathsetmacro{\r}{0.8}
\begin{tikzpicture}[baseline={([yshift=-.5ex]current bounding box.center)},every node/.style={font=\scriptsize}]
\draw [] (0,0) circle (\r cm);
\filldraw (\r,0) circle (1pt) node[right=0pt]{$5$};
\filldraw (72:\r) circle (1pt) node[above=0pt]{$4$};
\filldraw (144:\r) circle (1pt) node[above=0pt]{$3$};
\filldraw (-144:\r) circle (1pt) node[left=0pt]{$2$};
\filldraw (-72:\r) circle (1pt) node[below=0pt]{$1$};
\filldraw (\r/2,0) circle (1pt) (108:\r/2) circle (1pt) (-108:\r/2) circle (1pt);
\draw [thick] (\r,0) -- (\r/2,0) (72:\r) -- (108:\r/2) -- (144:\r) (-72:\r) -- (-108:\r/2) -- (-144:\r);
\draw [thick] (108:\r/2) -- (\r/2,0) node[pos=0.5,above right=-5pt]{};
\draw [thick] (-108:\r/2) -- (\r/2,0) node[pos=0.5,below right=-5pt]{};
\end{tikzpicture}&=\int \frac{dz_{a}d^{d}x_{a}^{\mu}}{z_{a}^{d+1}}\frac{dz_{b}d^{d}x_{b}^{\mu}}{z_{b}^{d+1}}\frac{dz_{c}d^{d}x_{c}^{\mu}}{z_{c}^{d+1}} K_{\Delta}^{\textrm{AdS}}(z_{a},x_{a};x_{1})K_{\Delta}^{\textrm{AdS}}(z_{a},x_{a};x_{2})  K_{\Delta}^{\textrm{AdS}}(z_{b},x_{b};x_{3}) \nonumber \\
&\times G_{\Delta}^{\textrm{AdS}}(z_{a},x_{a};z_{b},x_{b})G_{\Delta}^{\textrm{AdS}}(z_{b},x_{b};z_{c},x_{c})K_{\Delta}^{\textrm{AdS}}(z_{c},x_{c};x_{4})K_{\Delta}^{\textrm{AdS}}(z_{c},x_{c};x_{5})\ .
\end{align}
In these parameterizations, the number of integrals grows linearly with the number of vertices, and the integrand contains factors of ${}_{2}F_{1}$ functions, rendering even tree-level calculations rapidly intractable.

\subsection{de Sitter}
\label{desitterreview}

In dS, the calculations are very similar. The metric of the relevant patch of spacetime is given by 
\begin{equation}
ds^{2}=\frac{-d\eta^{2}+g_{\mu\nu}^{\textrm{E}}dx^{\mu}dx^{\nu}}{\eta^{2}}
\end{equation}
where $g_{\mu\nu}^{\textrm{E}}$ is the metric on $\mathbb{R}^{d}$. The $z$-coordinate has become conformal time, $\eta$. Conformal time is related to proper time, $t$, via 
\begin{equation}
\eta=e^{-t} 
\end{equation}
in dS units. The calculations follow those in AdS, differing only in the choice of propagators. We will consider the propagators necessary for computing correlation functions using the Schwinger-Keldysh in-in formalism in dS \cite{Schwinger:1960qe,Keldysh:1964ud}.\footnote{We discuss the computation of wavefunction coefficients in Section \ref{discussion}.}

The Wightman propagator in the Bunch-Davies vacuum \cite{Gibbons:1977mu,Bros:1994dn,Bros:1995js} is 
\begin{equation}\label{desitterprop}
G_{m}^{\textrm{dS}}(\eta,x;\eta',x')=\frac{\Gamma(\Delta_{+})\Gamma(\Delta_{-})}{(4\pi)^{\frac{d+1}{2}}\Gamma((d+1)/2)}{}_{2}F_{1}(\Delta_{+},\Delta_{-},\frac{d+1}{2},\chi)
\end{equation}
where 
\begin{equation}
\begin{split}
\chi&=\frac{(\eta+\eta')^{2}-(x-x')^{2}}{4\eta\eta'} \ , \\
\Delta_{\pm}&=\frac{1}{2}(d\pm \sqrt{d^{2}-4m^{2}}) \ .
\end{split}
\end{equation}
Note that $\Delta$ becomes complex for sufficiently heavy operators. 

The propagator in Eq. (\ref{desitterprop}) has a branch-cut when two points become time-like separated, which corresponds to $\chi \in (1,\infty)$. To go around the branch cut, we provide an $i\epsilon$ prescription:
\begin{equation}
\chi_{\pm}=\frac{(\eta+\eta')^{2}-(x-x')^{2}\pm i \textrm{sgn}(\eta-\eta')\epsilon}{4\eta\eta'}  \ ,
\end{equation}
where the choice of sign corresponds to two distinct operator orderings:\footnote{See Ref. \cite{Sleight:2019mgd} for a nice review.}
\begin{equation}\label{etimeprop}
\begin{split}
\langle \phi(\eta_{1},x_{1})\phi(\eta_{2},x_{2})\rangle&=G_{m}^{\textrm{dS}}(\chi_{+}) \ , \\
\langle \phi(\eta_{2},x_{2})\phi(\eta_{1},x_{1})\rangle&=G_{m}^{\textrm{dS}}(\chi_{-}) \ .
\end{split}
\end{equation}
The $i\epsilon$ prescription corresponding to the (anti-)time-ordered two-point correlator is
\begin{equation}\label{timeordered}
\chi_{\pm}^{T}=\frac{(\eta+\eta')^{2}-(x-x')^{2}\pm i \epsilon}{4\eta\eta'} \ ,
\end{equation}
such that 
\begin{equation}\label{timeprop}
\begin{split}
\langle T \phi(\eta_{1},x_{1})\phi(\eta_{2},x_{2})\rangle&=G_{m}^{\textrm{dS}}(\chi_{+}^{T}) \ , \\
\langle \bar{T} \phi(\eta_{1},x_{1})\phi(\eta_{2},x_{2})\rangle&=G_{m}^{\textrm{dS}}(\chi_{-}^{T}) \ . \\
\end{split}
\end{equation}
We study late time insertions where $\eta\rightarrow 0$, for which the (anti-)time-ordered propagator simplify to
\begin{equation}\label{latetimedS}
\begin{split}
\lim_{\eta \rightarrow 0}G_{m,\pm}^{\textrm{dS}}(\eta,x;\eta',x')&=\frac{2^{-d-1} \pi ^{-\frac{d}{2}-\frac{1}{2}} \Gamma \left(\Delta _+\right) \Gamma \left(\Delta _--\Delta _+\right)}{\Gamma \left(\frac{1}{2} \left(d-2 \Delta _++1\right)\right)}\left ( \frac{\eta \eta'}{(x-x')^{2}\mp i \epsilon-\eta'^{2}}\right )^{\Delta_{+}} \\
&+\frac{2^{-d-1} \pi ^{-\frac{d}{2}-\frac{1}{2}} \Gamma \left(\Delta _-\right) \Gamma \left(\Delta _+-\Delta _-\right)}{\Gamma \left(\frac{1}{2} \left(d-2 \Delta _-+1\right)\right)}\left ( \frac{\eta \eta'}{(x-x')^{2}\mp i \epsilon-\eta'^{2}}\right )^{\Delta_{-}} \ .
\end{split}
\end{equation}
Generally, we focus on computing the late-time behavior of either the $\Delta_{+}$ or $\Delta_{-}$ mode, which means picking one of the two terms in Eq. (\ref{latetimedS}), when including the late-time propagator. We will consider the $\eta^{\Delta_{+}}$ term for the remainder of the paper. 

To compute correlators, we can use the Schwinger-Keldysh in-in formalism~\cite{Schwinger:1960qe,Keldysh:1964ud}. This formalism requires including propagators with distinct $i\epsilon$ prescriptions from both Eqs.~(\ref{etimeprop}) and~(\ref{timeprop}) in a single integral. See Ref. \cite{Chen:2017ryl} for a nice review. However, the Witten diagrams are otherwise similar to those of AdS.

\section{The position space Witten-Feynman parameterization}\label{poswf}

We now compute the WF parameterization of a given Witten diagram in (A)dS. We begin with the Euclidean AdS case, where all integrals are manifestly convergent and no $i\epsilon$ prescription is required. We then treat the de Sitter case, which necessitates a contour deformation dictated by the $i\epsilon$ prescription. Next, we present combinatorial formulas for the $\mathcal{W}$ and $\mathcal{Y}$ polynomials as sums over subgraphs. We also show how conformal symmetry reduces the number of independent variables in an $n$-point Witten diagram to $n(n-3)/2$. We conclude with a brief discussion of how the WF parameterization is modified for BFSS, a non-conformal holographic theory.

\subsection{Witten-Feynman parameterization for anti-de Sitter}

We will restrict our kinematics to Euclidean AdS so that the integrals are manifestly convergent. The crucial input is the Schwinger form of the two-point propagators in Eqs. (\ref{adesitterprop}) and (\ref{eq:adsboundprop}):
\begin{equation}\label{scwhingerpre}
\begin{split}
G_{\Delta}^{\textrm{AdS}}(z,x^{\mu};z',x^{\mu}{}')=&z^{\Delta}z'^{\Delta}N^{\textrm{int}}_{\Delta}\int_{0}^{\infty} d\kappa  \int_{\kappa}^{\infty} d\beta  \ \beta^{d-1-\Delta}\\
&\times ( \kappa(\beta-\kappa))^{\Delta-\frac{d+1}{2}}e^{-(x-x')^{2}\beta-(z-z')^{2}\beta-4zz'\kappa}
\end{split}
\end{equation}
and
\begin{equation}\label{bulkboundaryprop}
K_{\Delta}^{\textrm{AdS}}(z,x;x')=z^{\Delta}N^{\textrm{ext}}_{\Delta}\int_{0}^{\infty} d\alpha \alpha^{\Delta-1}e^{-\alpha (z^{2}+(x-x')^{2})} \\ . 
\end{equation}
where the normalization factors are
\begin{equation}
\begin{split}
N^{\textrm{ext}}_{\Delta}&=\frac{\pi^{-d/2}}{\Gamma(\Delta-d/2)}  \ , \\
N^{\textrm{int}}_{\Delta}&=\frac{2^{2\Delta-d-1}}{\pi^{\frac{d+1}{2}}\Gamma(\Delta+\frac{1-d}{2})} \ .
\end{split}
\end{equation}
See Appendix \ref{dersch2f1} for a derivation of the Schwinger parameterizations of the two-point function (\ref{scwhingerpre}) from the propagator in Eq. (\ref{adesitterprop}). Upon inserting these forms of propagators into the Witten diagram, we can perform the integral over $x^{\mu}$ coordinates explicitly because they are Gaussian integrals. 

Before considering the general case, let us first consider the three-point Witten diagram as a simple example. The Feynman rules for the Witten diagram lead to the integral:
\begin{equation}
\pgfmathsetmacro{\r}{1}
\begin{tikzpicture}[baseline={([yshift=-.5ex]current bounding box.center)},every node/.style={font=\scriptsize}]
\filldraw (0,0) circle (1pt) ;
\draw [] (0,0) circle (\r cm);
\filldraw (0:\r) circle (1pt) node[right=0pt]{$3$};
\filldraw (120:\r) circle (1pt) node[above=0pt]{$2$};
\filldraw (240:\r) circle (1pt) node[below=0pt]{$1$};
\draw [thick] (0:\r) -- (0:0) ;
\draw [thick] (120:\r) -- (0:0) ;
\draw [thick] (240:\r) -- (0:0) ;
\end{tikzpicture} =W_{3}= \int \frac{dz d^{d}x}{z^{d+1}} K_{\Delta_{1}}(z,x;x_{1})K_{\Delta_{2}}(z,x;x_{2})K_{\Delta_{3}}(z,x;x_{3}) \ .
\end{equation}
We insert Eq. (\ref{bulkboundaryprop}) and then integrate over the bulk $x_{\mu}$. The result is 
\begin{equation}
\begin{split}
W_{3}=\pi^{d/2}N^{\textrm{ext}}_{\Delta_{1}}N^{\textrm{ext}}_{\Delta_{2}}N^{\textrm{ext}}_{\Delta_{3}} &\int \frac{d\alpha_{1}}{\alpha_{1}}\frac{d\alpha_{2}}{\alpha_{2}}\frac{d\alpha_{3}}{\alpha_{3}}\int dz \frac{\alpha_{1}^{\Delta_{1}}\alpha_{2}^{\Delta_{2}}\alpha_{3}^{\Delta_{3}}}{z^{d+1-\Delta_{1}-\Delta_{2}-\Delta_{3}}}\\
&\times \frac{e^{-\frac{x_{1,2}^{2}\alpha_{1}\alpha_{2}+x_{1,3}^{2}\alpha_{1}\alpha_{3}+x_{2,3}^{2}\alpha_{2}\alpha_{3}}{\alpha_{1}+\alpha_{2}+\alpha_{3}}-z^{2}(\alpha_{1}+\alpha_{2}+\alpha_{3})}}{(\alpha_{1}+\alpha_{2}+\alpha_{3})^{d/2}}
\end{split}
\end{equation}
from the Gaussian integration formula 
\begin{equation}
\int_{-\infty}^{\infty}d^{d}\vec{y}e^{-\vec{y}\cdot M\cdot \vec{y}+2\vec{w}\cdot \vec{y}}=\frac{\pi^{d/2}}{\det(M)}e^{\vec{w}\cdot M^{-1}\cdot \vec{w}} \ .
\end{equation}
We can now explicitly integrate out $z$, leaving the result 
\begin{equation}\label{schwinger3p}
\begin{split}
W_{3}=\frac{\pi^{d/2}}{2}\Gamma[\frac{1}{2}(\Delta_{123}-d)]N^{\textrm{ext}}_{\Delta_{1}}N^{\textrm{ext}}_{\Delta_{2}}N^{\textrm{ext}}_{\Delta_{3}} &\int \frac{d\alpha_{1}}{\alpha_{1}}\frac{d\alpha_{2}}{\alpha_{2}}\frac{d\alpha_{3}}{\alpha_{3}}\alpha_{1}^{\Delta_{1}}\alpha_{2}^{\Delta_{2}}\alpha_{3}^{\Delta_{3}} \frac{e^{-\frac{\mathcal{W}}{\mathcal{Y}}}}{\mathcal{Y}^{\frac{\Delta_{123}}{2}}}
\end{split}
\end{equation}
where 
\begin{equation}
\begin{split}
\Delta_{123}&=\Delta_{1}+\Delta_{2}+\Delta_{3} \ , \\
\mathcal{W}&=x_{1,2}^{2}\alpha_{1}\alpha_{2}+x_{1,3}^{2}\alpha_{1}\alpha_{3}+x_{2,3}^{2}\alpha_{2}\alpha_{3} \ , \\
\mathcal{Y}&=\alpha_{1}+\alpha_{2}+\alpha_{3} \ .
\end{split}
\end{equation}
We refer to parameterizations of the type in Eq.~(\ref{schwinger3p}), where the nontrivial part of the integrand appears in the exponential, along with their momentum-space analogs, as Schwinger-like parameterizations. We can convert Eq. (\ref{schwinger3p}) to a WF parameterization by inserting the delta function,
\begin{equation}
\forall \alpha_{i}>0:\quad 1=\int_{0}^{\infty} d\lambda \delta(1-\mathbf{m}_{1}\alpha_{1}-\mathbf{m}_{2}\alpha_{2}-\mathbf{m}_{3}\alpha_{3})
\end{equation}
re-scaling $\alpha_{i}\rightarrow \lambda \alpha_{i}$ and then integrating over $\lambda$. $\vec{\mathbf{m}}$ is a non-zero vector with non-negative coefficients. The result is 
\begin{equation}\label{WFthreepointtree}
\begin{split}
W_{3}=\frac{\pi^{d/2}}{2}\Gamma[\frac{1}{2}(\Delta_{123}-d)]\Gamma[\frac{\Delta_{123}}{2}]N^{\textrm{ext}}_{\Delta_{1}}N^{\textrm{ext}}_{\Delta_{2}}N^{\textrm{ext}}_{\Delta_{3}}&\int \frac{d\alpha_{1}}{\alpha_{1}}\frac{d\alpha_{2}}{\alpha_{2}}\frac{d\alpha_{3}}{\alpha_{3}}\alpha_{1}^{\Delta_{1}}\alpha_{2}^{\Delta_{2}}\alpha_{3}^{\Delta_{3}} \\
&\times \mathcal{W}^{\frac{-\Delta_{123}}{2}} \delta(1-\mathbf{m}_{1}\alpha_{1}-\mathbf{m}_{2}\alpha_{2}-\mathbf{m}_{3}\alpha_{3}) \ .
\end{split}
\end{equation}
This is the WF parameterization of the tree-level three-point Witten diagram. Note that it is trivial to evaluate the integral in this form. Choosing $\vec{\mathbf{m}}=(1,0,0)$ and then integrating over $\alpha_{2}$ and $\alpha_{3}$ sequentially, we find 
\begin{equation}
W=\frac{\pi^{d/2}\Gamma[\frac{1}{2}(\Delta_{123}-d)]\Gamma[\frac{\Delta_{12,3}}{2}]\Gamma[\frac{\Delta_{13,2}}{2}]\Gamma[\frac{\Delta_{23,1}}{2}]}{2 x_{12}^{\Delta_{12,3}}x_{13}^{\Delta_{13,2}}x_{23}^{\Delta_{23,1}}}
\end{equation}
where
\begin{equation}
\Delta_{ij,k}=\Delta_{i}+\Delta_{j}-\Delta_{k} \ .
\end{equation}
The tree-level three-point Witten diagram, largely fixed by conformal symmetry, has been known for decades. The benefit of the above computation strategy is that the WF parameterization in Eq. (\ref{WFthreepointtree}) generalizes to arbitrary Witten diagrams.

We now consider an arbitrary Witten diagram with any number of internal and external edges. We represent the exponent as 
\begin{equation}
\begin{split}
&\sum_{\{ e,(v)\}\in E^{\textrm{ext}}} \alpha_{e}((x_{i}-x_{v})^{2}+z_{v}^{2})+\sum_{\{ e,(v,v')\}\in E^{\textrm{int}}} \beta_{e}(x_{v}-x_{v'})^{2}+\beta_{e}(z_{v}-z_{v'})^{2}+4\kappa_{e}z_{v}z_{v'} \\
&=\sum_{v,v'\in V^{\textrm{int}}} x_{v}M_{v,v'}x_{v'}+\sum_{v\in V^{\textrm{int}}} 2 K_{v}\cdot x_{v}+J \ ,
\end{split}
\end{equation}
where $V^{\textrm{int}}$ denotes the set of internal vertices, $\{e, (v, v')\}$ denotes an internal edge along with the two internal vertices it connects, and $\{e, (v)\}$ denotes an external edge together with the internal vertex to which it is attached. We split $J$ into components, one independent of $z_{v}$ and one that depends on $z_{v}$,
\begin{equation}
J=J|_{z\rightarrow 0}+\mathcal{Z}
\end{equation}
where
\begin{equation}\label{genzformula}
\mathcal{Z}^{\textrm{AdS}}=\left [ \sum_{\{e,(v,v')\}\in E^{\textrm{ext}}} \beta_{e}(z_{v}-z_{v'})^{2}+4\kappa_{e} z_{v}z_{v'} \right ]+\left [\sum_{\{e,(v)\}\in E^{\textrm{ext}}} \alpha_{e}z_{v}^{2}\right ] \ .
\end{equation}
Importantly, removing the terms dependent on $z$ means that $J|_{z\rightarrow 0}$ depends only on $x_{i}$, $\beta_{e}$, and $\alpha_{e}$. The dependence on $\kappa_{e}$ and $z_{v}$ is entirely contained in $\mathcal{Z}$. Upon implementing the Gaussian integral, the Witten diagram becomes 
\begin{equation}\label{Wads}
\begin{split}
&W=\left ( \prod_{e\in E^{\textrm{ext}}}N^{\textrm{ext}}_{\Delta_{e}}\right ) \left ( \prod_{e\in E^{\textrm{int}}} N^{\textrm{int}}_{\Delta_{e}}\right )\pi^{d|V^{\textrm{int}}|/2}\int \left [ \prod_{\{e,(v)\}\in E^{\textrm{ext}}}d\alpha_{e} \alpha_{e}^{\Delta_{e}-1}z_{v}^{\Delta_{e}} \right ]\left [\prod_{v\in V^{\textrm{int}}} dz_{v}z_{v}^{-d-1}\right ]  \\
&\times \left [ \prod_{\{e,(v,v')\}\in E^{\textrm{int}}} d\beta_{e}d\kappa_{e} \beta_{e}^{d-1-\Delta_{e}}(\kappa_{e}(\beta_{e}-\kappa_{e}))^{\Delta_{e}-\frac{d+1}{2}}z_{v}^{\Delta_{e}}z_{v'}^{\Delta_{e}} \right ] \mathcal{Y}^{-d/2}e^{-\frac{\mathcal{W}}{\mathcal{Y}}-\mathcal{Z}^{\textrm{AdS}}}
\end{split}
\end{equation}
where the polynomials are given by 
\begin{equation}\label{deterform}
\begin{split}
\mathcal{Y}=\det(M), \quad \mathcal{W}=\det(M)(J|_{z_{v}\rightarrow 0}+K\cdot M^{-1}\cdot K) \ .
\end{split}
\end{equation}
Unlike the $x_{v}$ integrals, we cannot perform the $z$-integrals in the same manner because $z$ ranges from zero to infinity rather than over the entire real line. Consequently, we must treat the $z$ variables as genuine integration variables. The number of such variables is straightforward to count: each bulk-to-boundary propagator contributes one integration variable, each bulk-to-bulk propagator contributes two, and each bulk vertex contributes one. Thus, the total number of integration variables is  
\begin{equation}
\textrm{Number of Integration Variables} = 2|E^{\textrm{int}}| + |E^{\textrm{ext}}| + |V^{\textrm{int}}| \, .
\end{equation}
Finally, note that each $\beta_{e}$ integral runs from $\kappa_{e}$ to infinity.

We now convert the integral to the generalized Euler integral in Eq. (\ref{WFrepsketch}). We first insert the delta-function, 
\begin{equation}\label{delta1}
\forall z_{v}>0:\quad 1=\int_{0}^{\infty} d\lambda \ \delta(\lambda -\sum_{v} z_{v}) \ ,
\end{equation}
rescale all $z_{v}\rightarrow \lambda z_{v}$ and integrate over $\lambda$ using the identity. The integral over $\lambda$ pulls the $\mathcal{Z}$ polynomial down so that it is no longer in the exponent. We then insert the delta function,
\begin{equation}\label{delta2}
\forall \beta_{e},\kappa_{e},\alpha_{e}>0:\quad 1=\int d\lambda \delta(\lambda -\sum_{e} (\beta_{e}+\kappa_{e}) - \sum_{e} \alpha_{e} ) \ , 
\end{equation}
rescale all $\alpha_{e},\beta_{e},\kappa_{e}\rightarrow \lambda \alpha_{e},\lambda\beta_{e},\lambda\kappa_{e}$ and integrate over $\lambda$ again. The integration over $\lambda$ pulls down the factor of $\mathcal{W}/\mathcal{Y}$. The end result is the WF parameterization in Eq. (\ref{WFrepsketch}):
\begin{equation}\label{WFrepsketchfull}
\begin{split}
&W=K\int \left [ \prod_{\{e,{v}\}\in E^{\textrm{ext}}}d\alpha_{e} \alpha_{e}^{\Delta_{e}-1}z_{v}^{\Delta_{e}} \right ]\left [\prod_{v\in V^{\textrm{int}}} dz_{v}z_{v}^{-d-1}\right ]  \\
&\times \left [ \prod_{\{e,{v,v'}\}\in E^{\textrm{int}}} d\beta_{e}d\kappa_{e} \beta_{e}^{d-1-\Delta_{e}}(\kappa_{e}(\beta_{e}-\kappa_{e}))^{\Delta_{e}-\frac{d+1}{2}}z_{v}^{\Delta_{e}}z_{v'}^{\Delta_{e}} \right ] \frac{\mathcal{Z}^{\frac{d|V^{\textrm{int}}|}{2}-\Delta^{\textrm{int}}-\frac{\Delta^{\textrm{ext}}}{2}}}{\mathcal{Y}^{\frac{d-\Delta^{\textrm{ext}}}{2}}\mathcal{W}^{\frac{\Delta^{\textrm{ext}}}{2}}} \\
&\times \delta(1 -\sum_{e} (\beta_{e}+\kappa_{e}) - \sum_{e} \alpha_{e} )\delta(1 -\sum_{v} z_{v}) \ ,
\end{split}
\end{equation}
where 
\begin{equation}
\Delta^{\textrm{int}}=\sum_{e\in E^{\textrm{int}}}\Delta_{e}, \quad \Delta^{\textrm{ext}}=\sum_{e\in E^{\textrm{ext}}}\Delta_{e} \ ,
\end{equation}
and 
\begin{equation}
K=\frac{\pi^{d|V^{\textrm{int}}|/2}}{2}\left ( \prod_{e\in E^{\textrm{ext}}}N^{\textrm{ext}}_{\Delta_{e}}\right ) \left ( \prod_{e\in E^{\textrm{int}}} N^{\textrm{int}}_{\Delta_{e}}\right )\Gamma\left [\frac{\Delta^{\textrm{ext}}}{2}+\Delta^{\textrm{ext}} -\frac{d|V^{\textrm{int}}|}{2}\right ]\Gamma \left[ \frac{\Delta^{\textrm{ext}}}{2} \right ] \ .
\end{equation}
We note that the choice of delta functions in Eq. (\ref{WFrepsketchfull}) is arbitrary and reflects a particular choice of gauge for the two $GL(1)$ scaling redundancies. More concretely, we could replace both delta functions with 
\begin{equation}\label{gendeltafun}
\begin{split}
\delta(1-\sum z_{v})&\rightarrow \delta(1-\sum \mathbf{n}_{v}z_{v})\ , \\
\delta(1 -\sum_{e} (\beta_{e}+\kappa_{e}) - \sum_{e} \alpha_{e} )&\rightarrow\delta(1-\sum (\mathbf{m}_{\beta,e}\beta_{e}+\mathbf{m}_{\kappa,e}\kappa)-\sum \mathbf{m}_{\alpha,e}\alpha_{e})\ , \\
\end{split}
\end{equation}
where $\vec{\mathbf{m}}$ and $\vec{\mathbf{n}}$ are non-zero vectors with non-negative coefficients. However, we will stick to the choice of delta functions in Eq. (\ref{WFrepsketchfull}) for simplicity for the majority of the paper.  

\subsection{Witten-Feynman parameterization for de Sitter and contour deformations}\label{desitterwfpos}

We now turn to the WF parameterization in de Sitter space. We focus on Witten diagrams relevant for in-in correlators, using the propagators in Eqs.~(\ref{etimeprop}) and~(\ref{timeprop}). For simplicity, we will restrict to Witten diagrams where all propagators are $G_{m}^{\textrm{dS}}(\chi_{+}^{T})$. The other Witten diagrams are the same up to a $i\epsilon$ prescription. 

The derivation proceeds analogously to the AdS case, with the sole modification that we employ the time-ordered dS propagator:
\begin{equation}\label{dSGrep}
\begin{split}
G_{m}^{\textrm{dS}}(\eta,x;\eta',x')=N_{m}^{\textrm{int},\textrm{dS}}(\eta \eta')^{\Delta_{+}}&\int_{0}^{\infty} d\beta d\kappa \beta ^{\Delta _--1}(\kappa(\kappa+i\beta ))^{\Delta _+-\frac{d}{2}-\frac{1}{2}} \\
&\times e^{i\beta  \left((\eta -\eta')^2+i\epsilon-(x-x')^{2}\right)-4 \eta  \eta' \kappa }
\end{split}
\end{equation}
and use the bulk-to-boundary propagator
\begin{equation}\label{dSGrepbound}
\lim_{\eta\rightarrow 0}G_{m}^{\textrm{dS}}(\eta,x;\eta',x')|_{\eta^{\Delta_{+}}}=N_{m}^{\textrm{ext},\textrm{dS}}\eta'^{\Delta_{+}}\int_{0}^{\infty} d\alpha \alpha^{\Delta_{+}-1}e^{i\alpha(\eta'^{2}+i\epsilon-(x-x')^{2})}\ .
\end{equation}
where 
\begin{equation}
\begin{split}
N_{m}^{\textrm{int},\textrm{dS}}&=i^{\Delta_{-}} \frac{\pi ^{-\frac{d}{2}-\frac{1}{2}} 2^{2 \Delta _+-d-1}}{\Gamma \left(\frac{1}{2} \left(d-2 \Delta _-+1\right)\right)} \ , \\
N_{m}^{\textrm{ext},\textrm{dS}}&=i^{\Delta_{+}}\frac{\pi ^{\frac{1}{2} (-d-1)} 2^{2 \Delta _+-d-1} \Gamma \left(\Delta _--\Delta _+\right)}{\Gamma \left(\frac{d+1}{2}-\Delta _+\right)} \ . 
\end{split}
\end{equation}
To derive Eq. (\ref{dSGrep}), we used the Schwinger form of the ${}_{2}F_{1}$ function in Appendix \ref{dersch2f1} and then performed the Wick rotations $\alpha\rightarrow i\alpha$ $\beta\rightarrow i\beta$. Note that we do not Wick rotate $\kappa$. Due to the $i\epsilon$, the integrals are convergent in the large $\alpha$ and $\beta$ regime. Importantly, unlike the AdS propagator, the integral for $\beta$ is from zero to infinity. In principle, the derivation of the WF parameterized integral then proceeds analogously to the AdS case. We first obtain a Schwinger-like representation as in Eq.~(\ref{Wads}), and then convert it to a form resembling Eq.~(\ref{WFrepsketchfull}).\footnote{We are currently assuming that the integral is finite as we compute observables at arbitrarily late times, but there can generically be IR divergences.} The $i\epsilon$ can be absorbed in the $\mathcal{Z}$ polynomial:
\begin{equation}\label{Zds}
\mathcal{Z}^{\textrm{dS}}=\left [ \sum_{\{e,(v,v')\}\in E^{\textrm{ext}}}\beta_{e}((\eta_{v}-\eta_{v'})^{2}+i\epsilon) +i 4\kappa_{e} \eta_{v}\eta_{v'}\right ]+\left [\sum_{\{e,(v)\}\in E^{\textrm{ext}}} \alpha_{e}(\eta_{v}^{2}+i\epsilon)\right ] \ .
\end{equation}
Again, we are considering the case that all propagators are given by $G_{m}^{\textrm{dS}}(\chi_{+}^{T})$. We could similarly consider propagators with alternative $i\epsilon$ prescriptions, which would require some contour deformations of the form $\tilde{\beta}_{e}\rightarrow -\tilde{\beta}_{e}$ and $\tilde{\alpha}_{e}\rightarrow-\tilde{\alpha}_{e}$.

We could evaluate the resulting WF integral numerically at a fixed, small $\epsilon$. However, the integral becomes numerically unstable as we take $\epsilon$ smaller, making the prescription largely formal. We use a simpler, two-dimensional integral as an illustrative example of the problem, following Ref. \cite{Hannesdottir:2022bmo}. Consider the following integral
\begin{equation}\label{toymodel}
G(x,\epsilon)=I \int \frac{d\beta_{1}d\beta_{2}}{(\beta_{1}+\beta_{2})}e^{i\mathcal{V}-\epsilon(\beta_{1}+\beta_{2})}, \quad \mathcal{V}=\frac{x\beta_{1}\beta_{2}}{\beta_{1}+\beta_{2}}-(\beta_{1}+\beta_{2}) \ .
\end{equation}
In this integral, the $\epsilon$ is crucial for the large $\alpha_{1}$ and $\alpha_{2}$ integration regions to give a finite contribution. We can evaluate the integral analytically to 
\begin{equation}
G(x,\epsilon)=-\frac{4 i \tan ^{-1}\left(\frac{\sqrt{x}}{\sqrt{-x-4 i \epsilon +4}}\right)}{\sqrt{x} \sqrt{-x-4 i \epsilon +4}}\ .
\end{equation}
The $\epsilon$ specifies whether to pass above or below the branch point at $x=4$ for the $\epsilon=0$ function. We are ultimately interested in the limit $\epsilon \to 0^{+}$. For small but finite $\epsilon$, however, $G(x,\epsilon)$ clearly depends on $\epsilon$:
\begin{equation}
\frac{\partial G}{\partial \epsilon}(x,\epsilon)
= \frac{8x\,\arctan\!\left(\frac{\sqrt{x}}{\sqrt{4-x}}\right)}{\bigl[x(4-x)\bigr]^{3/2}}
- \frac{2}{x-4}
+ \mathcal{O}(\epsilon)\,, \qquad (\epsilon\to0^{+})\ ,
\end{equation}
We therefore need to take $\epsilon$ as small as possible in order to obtain an accurate result, since evaluating the integral at finite $\epsilon$ introduces an $\mathcal{O}(\epsilon)$ error. However, numerical integration gets progressively more unstable as we take $\epsilon \rightarrow 0$ if $x>4$. To address this issue, we seek a contour deformation that imposes
\begin{equation}
\textrm{Im}(\mathcal{V})>0 \ ,
\end{equation}
so we do not need an explicit $i\epsilon$. We use the deformation:
\begin{equation}\label{contourdefex}
\beta_{i}\rightarrow \beta_{i}+i\bar{\epsilon}\beta_{i}\partial_{\beta_{i}}\mathcal{V} \ . 
\end{equation}
At infinitesimal $\bar{\epsilon}$, this deforms the exponent as
\begin{equation}\label{deformedVex}
\hat{\mathcal{V}}=\mathcal{V}+i\bar{\epsilon}(\beta_{1}(\partial_{\beta_{1}}\mathcal{V})^{2}+\beta_{2}(\partial_{\beta_{2}}\mathcal{V})^{2})+\mathcal{O}(\bar{\epsilon}^{2}) \ .
\end{equation}
However, the important point is that $\textrm{Im}(\hat{\mathcal{V}})>0$ is actually valid for finite $\bar{\epsilon}$. For example, we can compute $\textrm{Im}(\hat{\mathcal{V}})$ explicitly and check that $\textrm{Im}(\hat{\mathcal{V}})>0$ for all $\beta_{1}$ and $\beta_{2}$ if $\bar{\epsilon}<\mathcal{O}(1)$ where the exact value depends on $x$. Therefore, we can set $\bar{\epsilon}=\mathcal{O}(1)$ and numerically evaluate the integral, finding that it converges rapidly. Note that there is also a Jacobian factor from the coordinate transformation in Eq. (\ref{contourdefex}).

We can apply the same computation strategy to the WF parameterized integrals in de Sitter. We again consider the case where all propagators are given by $G^{\textrm{dS}}(\chi^{T}_{+})$ for simplicity. We first write the integral in Schwinger form as 
\begin{align}\label{WformuladS}
&W=K^{\textrm{dS}} \left (\prod_{e\in E^{\textrm{ext}}}\eta^{\Delta_{e,+}}\right )\int \left [ \prod_{\{e,(v)\}\in E^{\textrm{ext}}}d\alpha_{e} \alpha_{e}^{\Delta_{+,e}-1}\eta_{v}^{\Delta_{e}} \right ]\left [\prod_{v\in V^{\textrm{int}}} d\eta_{v}\eta_{v}^{-d-1}\right ] \nonumber   \\
&\times \left [ \prod_{\{e,(v,v')\}\in E^{\textrm{int}}} d\beta_{e}d\kappa_{e} \beta_{e}^{\Delta_{e,-}-1}(\kappa_{e}(\kappa_{e}+i\beta_{e}))^{\Delta_{e,+}-\frac{d}{2}-\frac{1}{2}}\eta_{v}^{\Delta_{e,+}}\eta_{v'}^{\Delta_{e,+}} \right ] \mathcal{Y}^{-d/2} \ ,\\
&\times \exp\left [i\mathcal{V}-\sum_{\{e,(v,v')\}\in E^{\textrm{int}}}4\kappa_{e}\eta_{v}\eta_{v'}-\epsilon(\sum_{e\in E^{\textrm{int}}}\beta_{e}+\sum_{e\in E^{\textrm{ext}}}\alpha_{e}) \right ] \nonumber
\end{align}
where 
\begin{equation}
\mathcal{V}=\frac{\mathcal{W}}{\mathcal{Y}}+\mathcal{Z}|_{\kappa_{e}\rightarrow 0} \ ,
\end{equation}
and
\begin{equation}
K^{\textrm{dS}}=\left ( \prod_{e\in E^{\textrm{ext}}}N^{\textrm{ext},\textrm{dS}}_{\Delta_{e}}\right ) \left ( \prod_{e\in E^{\textrm{int}}} N^{\textrm{int},\textrm{dS}}_{\Delta_{e}}\right )\pi^{d|V^{\textrm{int}}|/2} \ .
\end{equation}
The purpose of the $i\epsilon$ is to ensure convergence of the integral at large $\beta_{e}$ and $\alpha_{e}$. Numerical convergence can again be achieved using a contour deformation. We consider the deformation 
\begin{equation}\label{contourde}
\begin{split}
\hat{\beta}_{e}&=\beta_{e}+i\bar{\epsilon} \beta_{e}\partial_{\beta_{e}}\mathcal{V} \ , \\
\hat{\alpha}_{e}&=\alpha_{e}+i\bar{\epsilon} \alpha_{e}\partial_{\alpha_{e}}\mathcal{V} \ , 
\end{split}
\end{equation}
where $\bar{\epsilon}$ is some numeric constant, so that 
\begin{equation}\label{deformedV}
\hat{\mathcal{V}}=\mathcal{V}+i\bar{\epsilon}(\sum_{e\in E^{\textrm{int}}}\beta_{e}(\partial_{\beta_{e}}\mathcal{V})^{2}+\sum_{e\in E^{\textrm{ext}}}\alpha_{e}(\partial_{\alpha_{e}}\mathcal{V})^{2}) +\mathcal{O}(\bar{\epsilon}^{2}) \ .
\end{equation}
The integral will give the same result as Eq. (\ref{WformuladS}) for any sufficiently small $\bar{\epsilon}$. Again, we do need to include the Jacobian factor for going from $\alpha_{e}$ and $\beta_{e}$ to $\hat{\alpha}_{e}$ and $\hat{\beta}_{e}$ in the numerical evaluation.

\subsection{Polynomials from subgraphs}

While the determinant parameterization is useful for numerical evaluation of $\mathcal{Y}$ and $\mathcal{Z}$, it obscures the combinatorial structure of these polynomials. In this section, we give a combinatorial formula for the $\mathcal{W}$ and $\mathcal{Y}$ polynomials as sums over subgraphs. We will only be interested in subgraphs that contain all internal vertices and have no cycles. Such subgraphs are not forests because they are allowed to exclude external vertices. We present the computational rules for $\mathcal{W}$ and $\mathcal{Y}$ and provide illustrative examples.

We first consider the $\mathcal{W}$ polynomial. The $\mathcal{W}$ polynomial is the only polynomial that depends on the external data, $x_{i}^{\mu}$. Furthermore, it only depends on the integration variables $\alpha_{e}$ and $\beta_{e}$, not $\kappa_{e}$ and $z_{v}$. It is defined as the sum over all subgraphs containing $(|V^{\textrm{int}}|+1)$ edges and all internal vertices. For any given sub-graph, there can only be a single connected component that links two external vertices, which determines the coefficient: $x_{i,j}^{2}$. For a given subgraph, the corresponding term is 
\begin{equation}
\mathcal{W}\ni x_{i,j}^{2}\prod_{e\in E^{\textrm{ext}}}\alpha_{e}\prod_{e\in E^{\textrm{int}}}\beta_{e} \ , 
\end{equation}
where $E^{\textrm{ext}}$ and $E^{\textrm{int}}$ are the sets of external and internal edges in the sub-graph. If the sub-graph does not have a connected component linking two external vertices, its coefficient is zero. For example, consider the four-point graph:
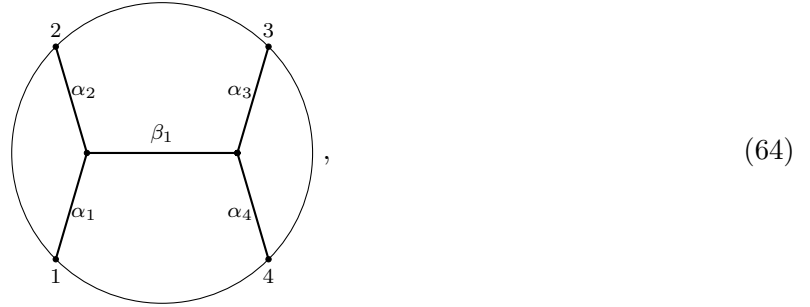
\begin{equation}\label{fourpoint}
\pgfmathsetmacro{\r}{2}
\begin{tikzpicture}[baseline={([yshift=-.5ex]current bounding box.center)},every node/.style={font=\scriptsize}]
\draw [] (0,0) circle (\r cm);
\filldraw (45:\r) circle (1pt) node[above=0pt]{$3$};
\filldraw (135:\r) circle (1pt) node[above=0pt]{$2$};
\filldraw (225:\r) circle (1pt) node[below=0pt]{$1$};
\filldraw (-45:\r) circle (1pt) node[below=0pt]{$4$};
\filldraw (-30:0.6*\r) circle (0pt) node[below=0pt]{$\alpha_{4}$};
\filldraw (30:0.6*\r) circle (0pt) node[above=0pt]{$\alpha_{3}$};
\filldraw (150:0.6*\r) circle (0pt) node[above=0pt]{$\alpha_{2}$};
\filldraw (210:0.6*\r) circle (0pt) node[below=0pt]{$\alpha_{1}$};
\filldraw (90:0*\r) circle (0pt) node[above=0pt]{$\beta_{1}$};
\filldraw (\r/2,0) circle (1pt) (0:\r/2) circle (1pt) (180:\r/2) circle (1pt);
\draw [thick] (45:\r) -- (0:\r/2) -- (-45:\r);
\draw [thick] (135:\r) -- (180:\r/2) -- (225:\r);
\draw [thick] (0:\r/2) -- (180:\r/2) ;
\end{tikzpicture} \ ,
\end{equation}
where we have labeled the edges with the relevant integration variables. Examples of subgraphs with a non-zero coefficient are 
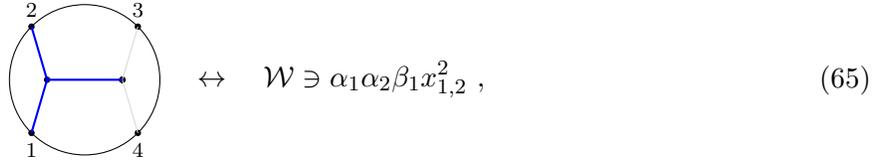
\begin{equation}
\begin{split}
\pgfmathsetmacro{\r}{1}
\begin{tikzpicture}[baseline={([yshift=-.5ex]current bounding box.center)},every node/.style={font=\scriptsize}]
\draw [] (0,0) circle (\r cm);
\filldraw (45:\r) circle (1pt) node[above=0pt]{$3$};
\filldraw (135:\r) circle (1pt) node[above=0pt]{$2$};
\filldraw (225:\r) circle (1pt) node[below=0pt]{$1$};
\filldraw (-45:\r) circle (1pt) node[below=0pt]{$4$};
\filldraw (\r/2,0) circle (1pt) (0:\r/2) circle (1pt) (180:\r/2) circle (1pt);
\draw [gray!30] (45:\r) -- (0:\r/2) -- (-45:\r);
\draw [blue,thick] (135:\r) -- (180:\r/2) -- (225:\r);
\draw [blue,thick] (0:\r/2) -- (180:\r/2) ;
\end{tikzpicture}  &\quad \leftrightarrow \quad \mathcal{W}\ni \alpha_{1}\alpha_{2}\beta_{1}x_{1,2}^{2}  \ , 
\end{split}
\end{equation}
\begin{equation}
\begin{split}
\pgfmathsetmacro{\r}{1}
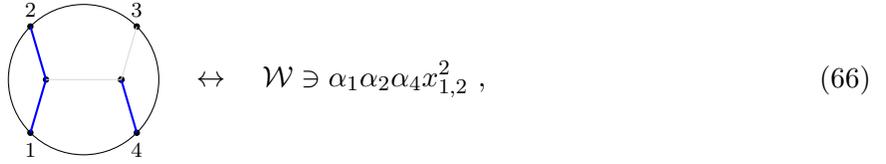
\begin{tikzpicture}[baseline={([yshift=-.5ex]current bounding box.center)},every node/.style={font=\scriptsize}]
\draw [] (0,0) circle (\r cm);
\filldraw (45:\r) circle (1pt) node[above=0pt]{$3$};
\filldraw (135:\r) circle (1pt) node[above=0pt]{$2$};
\filldraw (225:\r) circle (1pt) node[below=0pt]{$1$};
\filldraw (-45:\r) circle (1pt) node[below=0pt]{$4$};
\filldraw (\r/2,0) circle (1pt) (0:\r/2) circle (1pt) (180:\r/2) circle (1pt);
\draw [gray!30] (45:\r) -- (0:\r/2) ;
\draw [blue,thick] (0:\r/2) -- (-45:\r);
\draw [blue,thick] (135:\r) -- (180:\r/2) -- (225:\r);
\draw [gray!30] (0:\r/2) -- (180:\r/2) ;
\end{tikzpicture}  &\quad \leftrightarrow \quad \mathcal{W}\ni \alpha_{1}\alpha_{2}\alpha_{4}x_{1,2}^{2} \ ,
\end{split}
\end{equation}
\begin{equation}
\pgfmathsetmacro{\r}{1}
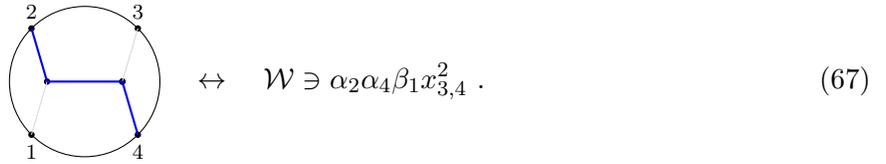
\begin{tikzpicture}[baseline={([yshift=-.5ex]current bounding box.center)},every node/.style={font=\scriptsize}]
\draw [] (0,0) circle (\r cm);
\filldraw (45:\r) circle (1pt) node[above=0pt]{$3$};
\filldraw (135:\r) circle (1pt) node[above=0pt]{$2$};
\filldraw (225:\r) circle (1pt) node[below=0pt]{$1$};
\filldraw (-45:\r) circle (1pt) node[below=0pt]{$4$};
\filldraw (\r/2,0) circle (1pt) (0:\r/2) circle (1pt) (180:\r/2) circle (1pt);
\draw [gray!30] (45:\r) -- (0:\r/2) ;
\draw [blue,thick] (0:\r/2) -- (-45:\r);
\draw [blue,thick] (135:\r) -- (180:\r/2);
\draw [gray!30] (180:\r/2) -- (225:\r);
\draw [blue,thick] (0:\r/2) -- (180:\r/2) ;
\end{tikzpicture}  \quad \leftrightarrow \quad \mathcal{W}\ni \alpha_{2}\alpha_{4}\beta_{1}x_{3,4}^{2}  \ .
\end{equation}
The blue highlight marks the subgraph linking the two external vertices. We reproduce the full polynomial here for convenience:
\begin{equation}\label{w4p}
\begin{split}
\mathcal{W}_{s}&=\alpha _1 \alpha _2 \beta _1 x_{1,2}^{2}+\alpha _1 \alpha _3 \beta _1 x_{1,3}^{2}+\alpha _1 \alpha _4 \beta _1 x_{1,4}^{2}+\alpha _1 \alpha _2 \alpha _3 x_{1,2}^{2}+\alpha _1 \alpha _2 \alpha _4 x_{1,2}^{2} \\
&+\alpha _2 \alpha _3 \beta _1 x_{2,3}^{2}+\alpha _2 \alpha _4 \beta _1 x_{2,4}^{2}+\alpha _3 \alpha _4 \beta _1 x_{3,4}^{2}+\alpha _1 \alpha _3 \alpha _4 x_{3,4}^{2}+\alpha _2 \alpha _3 \alpha _4 x_{3,4}^{2} \ .
\end{split}
\end{equation}
It is sufficiently simple that one can personally verify there are no missing terms. The $\mathcal{Y}$ polynomial is even simpler than the $\mathcal{W}$ polynomial because it does not depend on external kinematic data. It is defined as the sum over subgraphs with $|V^{\textrm{int}}|$ edges, where each connected component of the subgraph connects to at least one external vertex. The corresponding term is given by
\begin{equation}
\prod_{e\in E^{\textrm{ext}}}\alpha_{e}\prod_{e\in E^{\textrm{int}}}\beta_{e} \ .
\end{equation}
Using the four-point graph as an example again, two relevant subgraphs are
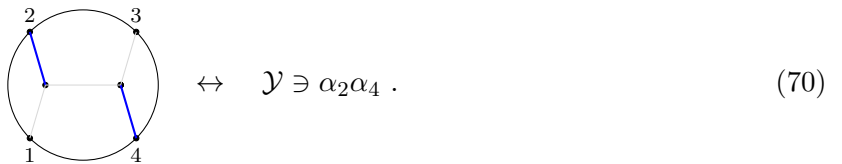
\begin{equation}
\pgfmathsetmacro{\r}{1}
\begin{tikzpicture}[baseline={([yshift=-.5ex]current bounding box.center)},every node/.style={font=\scriptsize}]
\draw [] (0,0) circle (\r cm);
\filldraw (45:\r) circle (1pt) node[above=0pt]{$3$};
\filldraw (135:\r) circle (1pt) node[above=0pt]{$2$};
\filldraw (225:\r) circle (1pt) node[below=0pt]{$1$};
\filldraw (-45:\r) circle (1pt) node[below=0pt]{$4$};
\filldraw (\r/2,0) circle (1pt) (0:\r/2) circle (1pt) (180:\r/2) circle (1pt);
\draw [gray!30] (45:\r) -- (0:\r/2) ;
\draw [blue,thick] (0:\r/2) -- (-45:\r);
\draw [blue,thick] (135:\r) -- (180:\r/2);
\draw [gray!30] (180:\r/2) -- (225:\r);
\draw [gray!30] (0:\r/2) -- (180:\r/2) ;
\end{tikzpicture}  \quad \leftrightarrow \quad \mathcal{Y}\ni \alpha_{2}\alpha_{4}  \ .
\end{equation}
\begin{equation}
\pgfmathsetmacro{\r}{1}
\begin{tikzpicture}[baseline={([yshift=-.5ex]current bounding box.center)},every node/.style={font=\scriptsize}]
\draw [] (0,0) circle (\r cm);
\filldraw (45:\r) circle (1pt) node[above=0pt]{$3$};
\filldraw (135:\r) circle (1pt) node[above=0pt]{$2$};
\filldraw (225:\r) circle (1pt) node[below=0pt]{$1$};
\filldraw (-45:\r) circle (1pt) node[below=0pt]{$4$};
\filldraw (\r/2,0) circle (1pt) (0:\r/2) circle (1pt) (180:\r/2) circle (1pt);
\draw [gray!30] (45:\r) -- (0:\r/2) ;
\draw [blue,thick] (0:\r/2) -- (-45:\r);
\draw [gray!30] (135:\r) -- (180:\r/2);
\draw [gray!30] (180:\r/2) -- (225:\r);
\draw [blue,thick] (0:\r/2) -- (180:\r/2) ;
\end{tikzpicture}  \quad \leftrightarrow \quad \mathcal{Y}\ni \beta_{1}\alpha_{4}  \ .
\end{equation}
The graph
\begin{equation}
\pgfmathsetmacro{\r}{1}
\begin{tikzpicture}[baseline={([yshift=-.5ex]current bounding box.center)},every node/.style={font=\scriptsize}]
\draw [] (0,0) circle (\r cm);
\filldraw (45:\r) circle (1pt) node[above=0pt]{$3$};
\filldraw (135:\r) circle (1pt) node[above=0pt]{$2$};
\filldraw (225:\r) circle (1pt) node[below=0pt]{$1$};
\filldraw (-45:\r) circle (1pt) node[below=0pt]{$4$};
\filldraw (\r/2,0) circle (1pt) (0:\r/2) circle (1pt) (180:\r/2) circle (1pt);
\draw [blue,thick] (45:\r) -- (0:\r/2) ;
\draw [blue,thick] (0:\r/2) -- (-45:\r);
\draw [gray!30] (135:\r) -- (180:\r/2);
\draw [gray!30] (180:\r/2) -- (225:\r);
\draw [gray!30] (0:\r/2) -- (180:\r/2) ;
\end{tikzpicture}  \quad \leftrightarrow \quad \mathcal{Y}\ni 0  \ .
\end{equation}
does not contribute to $\mathcal{Y}$ because it does not contain all vertices. We again reproduce the full polynomial here 
\begin{equation}\label{y4p}
\mathcal{Y}_{s}=\alpha _3 \beta _1+\alpha _1 \beta _1+\alpha _2 \beta _1+\alpha _4 \beta _1+\alpha _1 \alpha _3+\alpha _2 \alpha _3+\alpha _1 \alpha _4+\alpha _2 \alpha _4 \ .
\end{equation}
The $\mathcal{Z}$ polynomial is given in Eq. (\ref{schannelpoly})

We emphasize that the above structure holds beyond tree level. Consider the three-point triangle:
\begin{equation}
\label{trianglpicture}
\begin{tikzpicture}[baseline={([yshift=-.5ex]current bounding box.center)},every node/.style={font=\scriptsize}]\pgfmathsetmacro{\r}{2.2}
\draw [] (0,0) circle (\r cm);
\tikzset{decoration={snake,amplitude=.4mm,segment length=1.5mm,post length=0mm,pre length=0mm}}
\filldraw (0:\r) circle (1pt) node[right=0pt]{$1$};
\filldraw (120:\r) circle (1pt) node[above=0pt]{$2$};
\filldraw (240:\r) circle (1pt) node[below=0pt]{$3$};
\filldraw (0:\r/1.7) circle (1pt) node[right=0pt]{};
\filldraw (120:\r/1.7) circle (1pt) node[above=0pt]{};
\filldraw (240:\r/1.7) circle (1pt) node[below=0pt]{};
\filldraw (60:\r/2) circle (0pt) node{};
\filldraw (180:\r/1.6) circle (0pt) node{};
\filldraw (300:\r/2) circle (0pt) node{};
\draw [thick] (0:\r) -- (0:\r/1.7);
\draw [thick] (120:\r) -- (120:\r/1.7);
\draw [thick] (240:\r) -- (240:\r/1.7);
\draw [thick] (120:\r/1.7) -- (0:\r/1.7);
\draw [thick] (240:\r/1.7) -- (120:\r/1.7);
\draw [thick] (240:\r/1.7) -- (0:\r/1.7);
\filldraw (0:0.8*\r) circle (0pt) node[above=0pt]{$\alpha_{1}$};
\filldraw (120:0.85*\r) circle (0pt) node[right=0pt]{$\alpha_{2}$};
\filldraw (240:0.85*\r) circle (0pt) node[right=0pt]{$\alpha_{3}$};
\filldraw (60:0.4*\r) circle (0pt) node{$\beta_{1}$};
\filldraw (180:0.4*\r) circle (0pt) node{$\beta_{2}$};
\filldraw (300:0.4*\r) circle (0pt) node{$\beta_{3}$};
\end{tikzpicture} \ . 
\end{equation}
Some examples of contributing subgraphs to the $\mathcal{W}$ polynomial are 
\begin{equation}
\begin{tikzpicture}[baseline={([yshift=-.5ex]current bounding box.center)},every node/.style={font=\scriptsize}]\pgfmathsetmacro{\r}{1}
\draw [] (0,0) circle (\r cm);
\tikzset{decoration={snake,amplitude=.4mm,segment length=1.5mm,post length=0mm,pre length=0mm}}
\filldraw (0:\r) circle (1pt) node[right=0pt]{$1$};
\filldraw (120:\r) circle (1pt) node[above=0pt]{$2$};
\filldraw (240:\r) circle (1pt) node[below=0pt]{$3$};
\filldraw (0:\r/1.7) circle (1pt) node[right=0pt]{};
\filldraw (120:\r/1.7) circle (1pt) node[above=0pt]{};
\filldraw (240:\r/1.7) circle (1pt) node[below=0pt]{};
\filldraw (60:\r/2) circle (0pt) node{};
\filldraw (180:\r/1.6) circle (0pt) node{};
\filldraw (300:\r/2) circle (0pt) node{};
\draw [blue,thick] (0:\r) -- (0:\r/1.7);
\draw [gray!30] (120:\r) -- (120:\r/1.7);
\draw [blue,thick] (240:\r) -- (240:\r/1.7);
\draw [blue,thick] (120:\r/1.7) -- (0:\r/1.7);
\draw [gray!30] (240:\r/1.7) -- (120:\r/1.7);
\draw [blue,thick] (240:\r/1.7) -- (0:\r/1.7);
\end{tikzpicture}\quad  \leftrightarrow\quad  \mathcal{W}\ni \alpha_{3}\alpha_{1}\beta_{3}\beta_{1}(x_{1}-x_{3})^{2} \ ,
\end{equation}
\begin{equation}
\begin{tikzpicture}[baseline={([yshift=-.5ex]current bounding box.center)},every node/.style={font=\scriptsize}]\pgfmathsetmacro{\r}{1}
\draw [] (0,0) circle (\r cm);
\tikzset{decoration={snake,amplitude=.4mm,segment length=1.5mm,post length=0mm,pre length=0mm}}
\filldraw (0:\r) circle (1pt) node[right=0pt]{$1$};
\filldraw (120:\r) circle (1pt) node[above=0pt]{$2$};
\filldraw (240:\r) circle (1pt) node[below=0pt]{$3$};
\filldraw (0:\r/1.7) circle (1pt) node[right=0pt]{};
\filldraw (120:\r/1.7) circle (1pt) node[above=0pt]{};
\filldraw (240:\r/1.7) circle (1pt) node[below=0pt]{};
\filldraw (60:\r/2) circle (0pt) node{};
\filldraw (180:\r/1.6) circle (0pt) node{};
\filldraw (300:\r/2) circle (0pt) node{};
\draw [blue,thick] (0:\r) -- (0:\r/1.7);
\draw [gray!30] (120:\r) -- (120:\r/1.7);
\draw [blue,thick] (240:\r) -- (240:\r/1.7);
\draw [blue,thick] (120:\r/1.7) -- (0:\r/1.7);
\draw [blue,thick] (240:\r/1.7) -- (120:\r/1.7);
\draw [gray!30] (240:\r/1.7) -- (0:\r/1.7);
\end{tikzpicture}\quad  \leftrightarrow\quad  \mathcal{W}\ni \alpha_{3}\alpha_{1}\beta_{2}\beta_{1}(x_{1}-x_{3})^{2} \ .
\end{equation}
The full polynomial, $\mathcal{W}_{\textrm{tri}}$, is 
\begin{equation}
\begin{split}
\mathcal{W}_{\textrm{tri}}&=\alpha _1 \alpha _2 \alpha _3 \beta _1 x_{1,2}^2+\alpha _1 \alpha _2 \beta _1 \beta _2 x_{1,2}^2+\alpha _1 \alpha _2 \beta _1 \beta _3 x_{1,2}^2+\alpha _1 \alpha _2 \beta _2 \beta _3 x_{1,2}^2+\alpha _1 \alpha _3 \beta _1 \beta _2 x_{1,3}^2 \\
&+\alpha _1 \alpha _2 \alpha _3 \beta _3 x_{1,3}^2+\alpha _1 \alpha _3 \beta _1 \beta _3 x_{1,3}^2+\alpha _1 \alpha _3 \beta _2 \beta _3 x_{1,3}^2+\alpha _1 \alpha _2 \alpha _3 \beta _2 x_{2,3}^2+\alpha _2 \alpha _3 \beta _1 \beta _2 x_{2,3}^2\\
&+\alpha _2 \alpha _3 \beta _1 \beta _3 x_{2,3}^2+\alpha _2 \alpha _3 \beta _2 \beta _3 x_{2,3}^2\ .
\end{split}
\end{equation}
Some examples of contributing subgraphs to the $\mathcal{Y}$ polynomial are 
\begin{equation}
\begin{tikzpicture}[baseline={([yshift=-.5ex]current bounding box.center)},every node/.style={font=\scriptsize}]\pgfmathsetmacro{\r}{1}
\draw [] (0,0) circle (\r cm);
\tikzset{decoration={snake,amplitude=.4mm,segment length=1.5mm,post length=0mm,pre length=0mm}}
\filldraw (0:\r) circle (1pt) node[right=0pt]{$1$};
\filldraw (120:\r) circle (1pt) node[above=0pt]{$2$};
\filldraw (240:\r) circle (1pt) node[below=0pt]{$3$};
\filldraw (0:\r/1.7) circle (1pt) node[right=0pt]{};
\filldraw (120:\r/1.7) circle (1pt) node[above=0pt]{};
\filldraw (240:\r/1.7) circle (1pt) node[below=0pt]{};
\filldraw (60:\r/2) circle (0pt) node{};
\filldraw (180:\r/1.6) circle (0pt) node{};
\filldraw (300:\r/2) circle (0pt) node{};
\draw [blue,thick] (0:\r) -- (0:\r/1.7);
\draw [gray!30] (120:\r) -- (120:\r/1.7);
\draw [gray!30] (240:\r) -- (240:\r/1.7);
\draw [gray!30] (120:\r/1.7) -- (0:\r/1.7);
\draw [blue,thick] (240:\r/1.7) -- (120:\r/1.7);
\draw [blue,thick] (240:\r/1.7) -- (0:\r/1.7);
\end{tikzpicture}\quad  \leftrightarrow\quad  \mathcal{Y}\ni \alpha_{1}\beta_{3}\beta_{2} \ ,
\end{equation}
\begin{equation}
\begin{tikzpicture}[baseline={([yshift=-.5ex]current bounding box.center)},every node/.style={font=\scriptsize}]\pgfmathsetmacro{\r}{1}
\draw [] (0,0) circle (\r cm);
\tikzset{decoration={snake,amplitude=.4mm,segment length=1.5mm,post length=0mm,pre length=0mm}}
\filldraw (0:\r) circle (1pt) node[right=0pt]{$1$};
\filldraw (120:\r) circle (1pt) node[above=0pt]{$2$};
\filldraw (240:\r) circle (1pt) node[below=0pt]{$3$};
\filldraw (0:\r/1.7) circle (1pt) node[right=0pt]{};
\filldraw (120:\r/1.7) circle (1pt) node[above=0pt]{};
\filldraw (240:\r/1.7) circle (1pt) node[below=0pt]{};
\filldraw (60:\r/2) circle (0pt) node{};
\filldraw (180:\r/1.6) circle (0pt) node{};
\filldraw (300:\r/2) circle (0pt) node{};
\draw [blue,thick] (0:\r) -- (0:\r/1.7);
\draw [gray!30] (120:\r) -- (120:\r/1.7);
\draw [blue,thick] (240:\r) -- (240:\r/1.7);
\draw [gray!30] (120:\r/1.7) -- (0:\r/1.7);
\draw [gray!30] (240:\r/1.7) -- (120:\r/1.7);
\draw [blue,thick] (240:\r/1.7) -- (0:\r/1.7);
\end{tikzpicture}\quad  \leftrightarrow\quad  \mathcal{Y}\ni \alpha_{1}\beta_{3}\alpha_{3} \ .
\end{equation}
The full polynomial, $\mathcal{Y}_{\textrm{tri}}$, is 
\begin{equation}
\begin{split}
\mathcal{Y}_{\textrm{tri}}&=\alpha _1 \alpha _3 \beta _1+\alpha _2 \alpha _3 \beta _1+\alpha _1 \alpha _3 \beta _2+\alpha _3 \beta _1 \beta _2+\alpha _2 \alpha _3 \beta _3+\alpha _3 \beta _1 \beta _3+\alpha _3 \beta _2 \beta _3+\alpha _1 \alpha _2 \beta _2 \\
&+\alpha _1 \beta _1 \beta _2+\alpha _2 \beta _1 \beta _2+\alpha _1 \alpha _2 \beta _3+\alpha _1 \beta _1 \beta _3+\alpha _2 \beta _1 \beta _3+\alpha _1 \beta _2 \beta _3+\alpha _2 \beta _2 \beta _3+\alpha _1 \alpha _2 \alpha _3 \ .
\end{split}
\end{equation}

At higher loops, the subgraph must contain all internal vertices for the $\mathcal{W}$ polynomial. Consider the two-loop box 
\begin{equation}
\pgfmathsetmacro{\r}{2}
\begin{tikzpicture}[baseline={([yshift=-.5ex]current bounding box.center)},every node/.style={font=\scriptsize}]
\draw [] (0,0) circle (\r cm);
\filldraw (45:\r) circle (1pt) node[above=0pt]{$3$};
\filldraw (135:\r) circle (1pt) node[above=0pt]{$2$};
\filldraw (225:\r) circle (1pt) node[below=0pt]{$1$};
\filldraw (-45:\r) circle (1pt) node[below=0pt]{$4$};
\filldraw (45:\r/2) circle (1pt) node[above=0pt]{};
\filldraw (135:\r/2) circle (1pt) node[above=0pt]{};
\filldraw (225:\r/2) circle (1pt) node[below=0pt]{};
\filldraw (-45:\r/2) circle (1pt) node[below=0pt]{};
\draw [thick] (-45:\r/2) -- (225:\r/2) -- (135:\r/2) -- (45:\r/2) -- (-45:\r/2);
\draw [thick] (-45:\r/2) -- (-45:\r);
\draw [thick] (45:\r/2) -- (45:\r);
\draw [thick] (135:\r/2) -- (135:\r);
\draw [thick] (225:\r/2) -- (225:\r);
\draw [thick] (-135:\r/2) -- (45:\r/2);
\filldraw (0:0.45*\r) circle (0pt) node{$\beta_{3}$};
\filldraw (90:0.45*\r) circle (0pt) node{$\beta_{2}$};
\filldraw (180:0.45*\r) circle (0pt) node{$\beta_{1}$};
\filldraw (270:0.45*\r) circle (0pt) node{$\beta_{4}$};
\filldraw (45:0.85*\r) circle (0pt) node[left=0pt]{$\alpha_{3}$};
\filldraw (135:0.85*\r) circle (0pt) node[right=0pt]{$\alpha_{2}$};
\filldraw (-45:0.85*\r) circle (0pt) node[left=0pt]{$\alpha_{4}$};
\filldraw (-135:0.85*\r) circle (0pt) node[right=0pt]{$\alpha_{1}$};
\filldraw (135:0.2\r) circle (0pt) node{$\beta_{5}$};
\end{tikzpicture}  \ . 
\end{equation}
The following subgraphs do not contribute because they do not contain all internal vertices
\begin{equation}
\pgfmathsetmacro{\r}{1}
\begin{tikzpicture}[baseline={([yshift=-.5ex]current bounding box.center)},every node/.style={font=\scriptsize}]
\draw [] (0,0) circle (\r cm);
\filldraw (45:\r) circle (1pt) node[above=0pt]{$3$};
\filldraw (135:\r) circle (1pt) node[above=0pt]{$2$};
\filldraw (225:\r) circle (1pt) node[below=0pt]{$1$};
\filldraw (-45:\r) circle (1pt) node[below=0pt]{$4$};
\filldraw (45:\r/2) circle (1pt) node[above=0pt]{};
\filldraw (135:\r/2) circle (1pt) node[above=0pt]{};
\filldraw (225:\r/2) circle (1pt) node[below=0pt]{};
\filldraw (-45:\r/2) circle (1pt) node[below=0pt]{};
\draw [gray!30]  (225:\r/2) -- (135:\r/2) -- (45:\r/2);
\draw [blue,thick] (45:\r/2) -- (-45:\r/2) -- (-135:\r/2);
\draw [blue,thick] (-45:\r/2) -- (-45:\r);
\draw [gray!30] (45:\r/2) -- (45:\r);
\draw [gray!30] (135:\r/2) -- (135:\r);
\draw [blue,thick] (225:\r/2) -- (225:\r);
\draw [blue,thick] (-135:\r/2) -- (45:\r/2);
\end{tikzpicture} \quad \leftrightarrow \quad \mathcal{W}\ni 0 \ ,
\end{equation}
\begin{equation}
\pgfmathsetmacro{\r}{1}
\begin{tikzpicture}[baseline={([yshift=-.5ex]current bounding box.center)},every node/.style={font=\scriptsize}]
\draw [] (0,0) circle (\r cm);
\filldraw (45:\r) circle (1pt) node[above=0pt]{$3$};
\filldraw (135:\r) circle (1pt) node[above=0pt]{$2$};
\filldraw (225:\r) circle (1pt) node[below=0pt]{$1$};
\filldraw (-45:\r) circle (1pt) node[below=0pt]{$4$};
\filldraw (45:\r/2) circle (1pt) node[above=0pt]{};
\filldraw (135:\r/2) circle (1pt) node[above=0pt]{};
\filldraw (225:\r/2) circle (1pt) node[below=0pt]{};
\filldraw (-45:\r/2) circle (1pt) node[below=0pt]{};
\draw [gray!30]  (225:\r/2) -- (135:\r/2) -- (45:\r/2);
\draw [blue,thick] (45:\r/2) -- (-45:\r/2) -- (-135:\r/2);
\draw [blue,thick] (-45:\r/2) -- (-45:\r);
\draw [blue,thick] (45:\r/2) -- (45:\r);
\draw [gray!30] (135:\r/2) -- (135:\r);
\draw [blue,thick] (225:\r/2) -- (225:\r);
\draw [gray!30] (-135:\r/2) -- (45:\r/2);
\end{tikzpicture} \quad \leftrightarrow \quad \mathcal{W}\ni 0 \ .
\end{equation}
We do not reproduce the full polynomials for this graph in text due to their length.

\subsection{Reducing independent variables via conformal symmetry}\label{levconf}

Although the position and momentum WF parameterizations have nice combinatorial structure, they are functions of many variables and, therefore, difficult to compute. In this section, we discuss how we can leverage bulk isometries to simplify the Witten diagrams and (partially) alleviate the issue. The bulk isometries of (A)dS become conformal symmetries of the boundary correlators, which are given by 
\begin{equation}
\begin{split}
P_{\mu,i}&=-i\partial_{x_{i}^{\mu}}\ , \quad D_{i}=-i\sum_{\mu=1}^{d}x_{i}^{\mu}\partial_{x_{i}^{\mu}} \ , \\
L_{\mu\nu,i}&=i(x_{i,\mu}\partial_{x_{i}^{\nu}}-x_{\nu}\partial_{x_{i}^{\mu}})\ , \quad K_{\mu,i}=i(x_{i}^{2}\partial_{x_{i}^{\mu}}-i2x_{i,\mu}D)\ ,
\end{split}
\end{equation}
where $x_{i}^{\mu}$ is the location of the $i$th operator insertion \cite{DiFrancesco:1997nk}. Bulk isometry imposes that 
\begin{equation}
\begin{split}
0&=\left ( \sum_{i=1}^{n}P_{\mu,i}\right )W\ , \quad 0=\left ( \sum_{i=1}^{n}D\right )W \ , \\
0&=\left ( \sum_{i=1}^{n}L_{\mu\nu,i}\right )W\ , \quad 0=\left ( \sum_{i=1}^{n}K_{\mu,i}\right )W \ ,
\end{split}
\end{equation}
for any individual conformal Witten diagram.

Conformal symmetry places strong constraints on the kinematic dependence of generic Witten diagrams. For example, conformal symmetry is sufficient to completely fix the kinematic dependence of any three-point Witten diagram to the form 
\begin{equation}
\langle \mathcal{O}_{\Delta_{1}}(x_{1})\mathcal{O}_{\Delta_{2}}(x_{2})\mathcal{O}_{\Delta_{3}}(x_{3})\rangle=\frac{C}{x_{12}^{\Delta_{1}+\Delta_{2}-\Delta_{3}}x_{13}^{\Delta_{1}+\Delta_{3}-\Delta_{2}}x_{23}^{\Delta_{2}+\Delta_{3}-\Delta_{1}}} \ , 
\end{equation}
where $C$ is some numeric constant that is some function of $\Delta_{i}$ and $d$. At four-point, bulk isometries are not sufficient to uniquely fix the kinematic dependence, but still constrain any four-point Witten diagram to the form
\begin{equation}
\langle \mathcal{O}_{\Delta_{1}}(x_{1})\mathcal{O}_{\Delta_{2}}(x_{2})\mathcal{O}_{\Delta_{3}}(x_{3})\mathcal{O}_{\Delta_{4}}(x_{4})\rangle=f(\eta,\xi) \prod_{i<j}x_{ij}^{\frac{\sum_{i=1}^{4} \Delta_{i}}{3}-\Delta_{i}-\Delta_{j}}
\end{equation}
where the conformal cross-ratios are
\begin{equation}
\eta=\frac{x_{13}^{2}x_{24}^{2}}{x_{12}^{2}x_{34}^{2}}, \quad \xi=\frac{x_{23}^{2}x_{14}^{2}}{x_{12}^{2}x_{34}^{2}} \ .
\end{equation}
See Ref. \cite{DiFrancesco:1997nk} for a derivation. A general $n$-point Witten diagram is a non-trivial function of $n(n-3)/2$ cross ratios \cite{DiFrancesco:1997nk}. 

We restrict ourselves to four-point. To leverage the above symmetry constraints, we work with a special choice of kinematics without loss of generality. We restrict to 
\begin{equation}
x_{1}^{\mu}=0, \quad x_{2}^{\mu}=e^{\mu}, \quad x_{4}^{\mu}\rightarrow \infty
\end{equation}
where $e^{\mu}$ is some arbitrary constant vector such that $e^{2}=1$. We then find the cross-ratios reduce to 
\begin{equation}\label{etaxi}
\eta=x_{3}^{2}, \quad \xi=(e-x_{3})^{2}
\end{equation}
and that 
\begin{equation}\label{asymptoticlimitx4}
\begin{split}
\lim_{x_{4}\rightarrow \infty}\langle \mathcal{O}_{\Delta_{1}}(0)\mathcal{O}_{\Delta_{2}}(e^{\mu})\mathcal{O}_{\Delta_{3}}(x_{3})\mathcal{O}_{\Delta_{4}}(x_{4})\rangle=\frac{\eta^{\frac{1}{6}(\Delta_{4}+\Delta_{2}-2(\Delta_{1}+\Delta_{3}))}\xi^{\frac{1}{6}(\Delta_{1}+\Delta_{4}-2(\Delta_{2}+\Delta_{3}))}}{x_{4}^{-2\Delta_{4}}}f(\eta,\xi) 
\end{split}
\end{equation}
Our goal is to compute $F(\eta,\xi)$ using the WF parameterizations, which requires taking the limit that $x_{4}\rightarrow \infty$. In principle, it is actually quite non-trivial to compute integrals in the limit that particular parameters become large. Fortunately, there are well-established computational strategies for evaluating generalized Euler integrals in the limit where one of the variables becomes large, collectively known as the method of regions \cite{Smirnov:1991jn, Beneke:1997zp, Smirnov:1998vk, Smirnov:1999bza, Pak:2010pt, Ananthanarayan:2018tog, Ananthanarayan:2020ptw}.\footnote{See Appendix B of Ref. \cite{Biggs:2025qfh} for a quick introduction.}

Applying the method of regions to the WF representation in position space, we conjecture that the leading contribution as $x_{4}\!\to\!\infty$ is captured by the rescaling
\begin{equation}\label{usubmr}
\alpha_{4}\ \to\ \frac{\alpha_{4}}{x_{4}^{2}} \, ,
\end{equation}
followed by taking $|x_{4}|\!\to\!\infty$ in the integrand (with the remaining integration variables assumed to be $\mathcal{O}(1)$). 
This procedure isolates the factor $x_{4}^{-2\Delta_{4}}$ 
appearing in the asymptotic limit (\ref{asymptoticlimitx4}). 
The resulting integral depends only on the cross-ratios $\eta$ and $\xi$, 
from which $f(\eta,\xi)$ can be extracted using Eqs.~(\ref{etaxi}) and~(\ref{asymptoticlimitx4}). We emphasize that taking a series expansion of the integrand as above and integrating do not commute. For example, directly taking the $|x_{4}|\rightarrow \infty$ limit of the integrand without performing the $u$-substitution in Eq.~(\ref{usubmr}) yields a different result upon integration. The above procedure isolates the leading contribution of a single contributing region in the $x_{4}\rightarrow \infty$ limit. We have conjectured, but not proven, that all other regions are subdominant or do not exist for all conformal Witten diagrams. One can explicitly check this conjecture for any individual Witten diagram. 

As an illustrative example, consider the WF parameterization of the tree-level three-point Witten diagram in Eq.~(\ref{WFthreepointtree}) and consider the $x_{3}\rightarrow \infty$ limit of the integral. Applying the above procedure, we obtain
\begin{equation}
\begin{split}
\lim_{x_{3}\rightarrow \infty} W_{3} 
&= \frac{1}{x_{3}^{2\Delta_{3}}}\,\frac{\pi^{d/2}}{2}\,
   \Gamma\!\left[\tfrac{1}{2}(\Delta_{123}-d)\right]
   \Gamma\!\left[\tfrac{\Delta_{123}}{2}\right]\,
   N^{\mathrm{ext}}_{\Delta_{1}} N^{\mathrm{ext}}_{\Delta_{2}} N^{\mathrm{ext}}_{\Delta_{3}}\int \frac{d\alpha_{1}}{\alpha_{1}}
                 \frac{d\alpha_{2}}{\alpha_{2}}
                 \frac{d\alpha_{3}}{\alpha_{3}} \\
&\quad \times
                 \alpha_{1}^{\Delta_{1}}
                 \alpha_{2}^{\Delta_{2}}
                 \alpha_{3}^{\Delta_{3}}
                 \big(\alpha_{1}\alpha_{2}x_{12}^{2}
                     +\alpha_{1}\alpha_{3}
                     +\alpha_{2}\alpha_{3}\big)^{-\tfrac{\Delta_{123}}{2}}
                 \delta(1-\mathbf{m}_{1}\alpha_{1}-\mathbf{m}_{2}\alpha_{2}) \, .
\end{split}
\end{equation}
This integral can be evaluated exactly, yielding
\begin{equation}
\lim_{x_{3}\rightarrow \infty} W_{3} 
= \frac{\pi^{d/2}\,
   \Gamma\!\left[\tfrac{1}{2}(\Delta_{123}-d)\right]
   \Gamma\!\left[\tfrac{\Delta_{12,3}}{2}\right]
   \Gamma\!\left[\tfrac{\Delta_{13,2}}{2}\right]
   \Gamma\!\left[\tfrac{\Delta_{23,1}}{2}\right]}
   {2\,x_{12}^{\Delta_{12,3}}\,x_{3}^{2\Delta_{3}}} \, ,
\end{equation}
which precisely matches the result obtained by first evaluating the original integral in Eq.~(\ref{WFthreepointtree}) and then taking the limit.

\subsection{An aside on BFSS}\label{asideBFSS}

In this subsection, we briefly comment on the potential applicability of the WF parameterization to the BFSS matrix model \cite{deWit:1988wri,Banks:1996vh,Susskind:1997cw,Seiberg:1997ad,Sen:1997we,Itzhaki:1998dd,Polchinski:1999br}. The BFSS matrix model is a \textit{non-conformal} quantum mechanical model that is the dimensional reduction of the $\mathcal{N}=4$ super-Yang-Mills Lagrangian to one dimension. At strong coupling, the holographic dual of BFSS is Type IIA supergravity on a Weyl rescaling of  AdS${}_{2}\times \textrm{S}_{8}$
\begin{equation}\label{metricintro}
ds^{2}\propto z^{3/5}\left [ \left ( \frac{2}{5}\right )^{2}\left (\frac{d\tau^{2}+dz^{2}}{z^{2}}\right )+d\Omega_{8}^{2}  \right ]  \ .
\end{equation}
See Refs.~\cite{Itzhaki:1998dd,Biggs:2025qfh} for discussions of the classes of observables that can be reliably computed within the framework of this holographic duality. From the perspective of experimental comparison, higher-point computations in the BFSS model are particularly urgent, as a quantum computer with around 7000 qubits could conjecturally access the relevant holographic regime \cite{Maldacena:2023acv}.\footnote{Another holographic regime arises at parametrically low temperatures and long times, where the dual description is eleven-dimensional supergravity on $\mathbb{R}^{1,10}$ \cite{Polchinski:1999br}; see Refs.~\cite{Herderschee:2023pza,Herderschee:2023bnc} for recent results.}

The Weyl rescaling breaks the AdS${}_{2}$ bulk isometry group, reflecting the fact that BFSS is not conformal. One might then ask whether the formalism developed in this paper still applies. Fortunately, Ref.~\cite{Biggs:2025qfh} shows that the bulk-to-bulk and bulk-to-boundary propagators are proportional to those of AdS${}_{2}$, differing only by a trivial factor:
\begin{equation}\label{dimensionrel}
\begin{split}
K^{\textrm{BFSS}}_{\Delta}(z,\tau;\tau')&\propto z^{9/10}K^{\textrm{AdS}_{2}}_{\Delta-9/10}(z,\tau;\tau') \\
G^{\textrm{BFSS}}_{\Delta}(z,\tau;z',\tau')&\propto (zz')^{9/10}G^{\textrm{AdS}_{2}}_{\Delta-9/10}(z,\tau;z',\tau') \ . 
\end{split}
\end{equation}
As a result, the WF parameterization of the Witten diagrams relevant to BFSS coincides with that of conformal holographic theories, apart from modifications to the exponents. In addition, the conformal constraints discussed in the previous section no longer apply, so the $n$-point Witten diagrams depend on $n(n-1)/2-1$ independent cross-ratios. We will examine the tree-level three-point correlator in Section~\ref{bfss3p}.

\section{The momentum space Witten-Feynman parameterization}\label{wittenrevmom}

We now consider the momentum space parameterizations of boundary correlators. We consider only Euclidean AdS for simplicity. The generalization to dS proceeds in the same manner as Section \ref{desitterwfpos} and is not done out explicitly. We discuss the combinatorial formulas for the relevant polynomials in Section \ref{polynomialsfromforests}. 

\subsection{Witten-Feynman parameterization for anti-de Sitter}

We define momentum space correlation functions via a Fourier transform:
\begin{equation}
\langle \tilde{\mathcal{O}}(k_{1})\tilde{\mathcal{O}}(k_{2})\ldots \tilde{\mathcal{O}}(k_{n})\rangle= \int \left [ \prod_{i} \frac{d^{d}x_{i}}{(2\pi)^{d}} e^{-ik_{i}\cdot x_{i}} \right] \langle \mathcal{O}(x_{1})\mathcal{O}(x_{2})\ldots \mathcal{O}(x_{n})\rangle \ .
\end{equation}
Translation invariance of any correlation function implies that its Fourier transform has support only on a delta function:
\begin{equation}
\langle \tilde{\mathcal{O}}(k_{1})\tilde{\mathcal{O}}(k_{2})\ldots \tilde{\mathcal{O}}(k_{n})\rangle= \delta^{d}(\sum_{i}k_{i})\langle\langle \tilde{\mathcal{O}}(k_{1})\tilde{\mathcal{O}}(k_{2})\ldots \tilde{\mathcal{O}}(k_{n})\rangle \rangle \ ,
\end{equation}
where we use double brackets to denote the delta-function stripped momentum-space correlation function. To isolate the delta-function stripped correlation function, we can consider 
\begin{equation}
\begin{split}
\langle\langle \tilde{\mathcal{O}}(k_{1})\tilde{\mathcal{O}}(k_{2})\ldots \tilde{\mathcal{O}}(k_{n})\rangle \rangle &=\int d^{d}k_{1} \langle \tilde{\mathcal{O}}(k_{1})\tilde{\mathcal{O}}(k_{2})\ldots \tilde{\mathcal{O}}(k_{n})\rangle \\
&=\int \left [ \prod_{i>2} \frac{d^{d}x_{i}}{(2\pi)^{d}} e^{-ik_{i}\cdot x_{i}} \right] \langle \mathcal{O}(0)\mathcal{O}(x_{2})\ldots \mathcal{O}(x_{n})\rangle \ ,
\end{split}
\end{equation}
where we have ``gauge fixed" the translation symmetry.

Converting to momentum space primarily benefits the calculation by simplifying the bulk integration over $x_{v}^{\mu}$. The momentum space propagators, 
\begin{equation}
\begin{split}
K_{\Delta}^{\textrm{AdS}}(z,x;x')&=\int \frac{d^{d} x}{(2\pi)^{d}}e^{ik\cdot(x- x')}\tilde{K}_{\Delta}^{\textrm{AdS}}(z,k), \\ G^{\textrm{AdS}}_{\Delta}(z,x;z',x')&=\int \frac{d^{d} x}{(2\pi)^{d}}e^{ik\cdot(x- x')} \tilde{G}^{\textrm{AdS}}_{\Delta}(z,z',k) \ ,
\end{split}
\end{equation}
are
\begin{equation}\label{momentbound}
\tilde{K}_{\Delta}^{\textrm{AdS}}(2z,k)=\tilde{N}^{\textrm{ext}}_{\Delta}\int_{0}^{\infty} d\tilde{\alpha} \tilde{\alpha}^{d/2-\Delta-1} e^{-\frac{z^{2}}{4\tilde{\alpha}}-\tilde{\alpha}k^{2}} \ ,
\end{equation}
and 
\begin{equation}\label{momentbulk}
\begin{split}
\tilde{G}^{\textrm{AdS}}_{\Delta}(2z,2z',k)&=z^{\Delta}z'^{\Delta}\tilde{N}^{\textrm{int}}_{\Delta} \int_{0}^{\infty} d\tilde{\kappa}\int_{0}^{\tilde{\kappa}}d\tilde{\beta} \tilde{\kappa}^{d-1-2\Delta}\tilde{\beta}^{\frac{-1}{2}}(\tilde{\kappa}-\tilde{\beta})^{\Delta-\frac{d+1}{2}}e^{-\tilde{\beta}k^{2}-\frac{(z-z')^{2}}{\tilde{\beta}}-\frac{4zz'}{\tilde{\kappa}}} \ .
\end{split}
\end{equation}
The normalization factors are 
\begin{equation}
\begin{split}
\tilde{N}^{\textrm{int}}_{\Delta}&=N^{\textrm{int}}_{\Delta} 2^{d}\pi^{d/2} \ , \\
\tilde{N}^{\textrm{ext}}_{\Delta}&=N^{\textrm{ext}}_{\Delta} 2^{d-\Delta}\pi^{d/2} \ .
\end{split}
\end{equation}
Although we still need to integrate over the $z_{v}$ of each bulk vertex, working in momentum space trivializes many of the bulk $k_{v}^{\mu}$ integrals due to momentum conservation. Note that the additional factors of two in the $z$ and $z'$ in Eqs. (\ref{momentbound}) and (\ref{momentbulk}) can be removed via a u-sub $z_{v}\rightarrow z_{v}/2$ in the integral, which leads to a Jacobian factor of 
\begin{equation}
\mathcal{J}_{z\rightarrow z/2}=2^{d|V^{\textrm{int}}|-\Delta^{\textrm{ext}}-2\Delta^{\textrm{int}}}.
\end{equation}

Upon inserting Eqs. (\ref{momentbound}) and (\ref{momentbulk}) into the momentum space Witten diagram, we find an exponent of the form  
\begin{equation}
\begin{split}
&\sum_{e\in E^{\textrm{ext}}} (\tilde{\alpha}_{e}k_{e}^{2}+\frac{z_{v}^{2}}{\tilde{\alpha}_{e}})+\sum_{e\in E^{\textrm{int}}} \tilde{\beta}_{e}k_{e}^{2}+\frac{1}{\tilde{\beta}_{e}}(z_{v}-z_{v'})^{2}+\frac{4z_{v}z_{v'}}{\tilde{\kappa}_{e}} \\
&=\sum_{r,s=1}^{L} \ell_{r}M_{r,s}\ell_{s}+\sum_{s}^{L} 2 K_{s}\cdot \ell_{s}+J \ ,
\end{split}
\end{equation}
where $L$ is the loop-order of the Witten diagram, and $k_{e}$ is the momentum of edge $e$ after imposing momentum conservation. We again break up $J$ into two components independent of $z_{v}$ and not independent of $z_{v}$:
\begin{equation}
J=J|_{z_{v}\rightarrow 0}+\tilde{\mathcal{Z}}^{\textrm{AdS}} \ ,
\end{equation}
where 
\begin{equation}
\tilde{\mathcal{Z}}^{\textrm{AdS}}=\mathcal{Z}^{\textrm{AdS}}|_{\alpha_{e}\rightarrow (\tilde{\alpha_{e}})^{-1},\beta_{e}\rightarrow (\tilde{\beta_{e}})^{-1},\kappa_{e}\rightarrow (\tilde{\kappa_{e}})^{-1}} \ .
\end{equation}
Importantly, the part of the exponent independent of $\tilde{\kappa}_{e}$, $z_{v}$ and $\tilde{\alpha}_{e}$ is \textit{identical} to the exponent one would derive in a flat space amplitude with massless internal states and off-shell external momentum. The momentum space analog of the $\mathcal{Y}$ and $\mathcal{Z}$ integral are therefore 
\begin{equation}\label{momentumpoly}
\tilde{\mathcal{Y}}=\mathcal{U}(\tilde{\beta}_{e}), \quad \tilde{\mathcal{W}}=\mathcal{F}(k_i,\tilde{\beta}_{e})+\mathcal{U}(\tilde{\beta}_{e})\times \left ( \sum_{e\in E^{\textrm{ext}}} \tilde{\alpha}_{e}k^{2}_{e} \right )\ , 
\end{equation}
where $\mathcal{U}(\tilde{\beta}_{e})$ and $\mathcal{F}(k_i,\tilde{\beta}_{e})$ are the first and second Symanzik polynomials that appear in the standard Feynman parameterization of Feynman integrals and only functions of $\tilde{\beta}_{e}$ and $k_{i}^{\mu}$. The term in brackets in Eq. (\ref{momentumpoly}) is the polynomial $\mathcal{S}$ referenced in Eq. (\ref{WFrepsketch}) of the introduction. 

Using the above polynomials, we express the momentum-space Witten diagram in WF parameterization by integrating over the loop momentum, resulting in an integral in a Schwinger-like form. We can then insert the delta functions and perform u-substitution as described in  Eqs. (\ref{delta1}) and (\ref{delta2}) to get the result in WF parameterization:
\begin{equation}\label{swchingerep}
\begin{split}
&\tilde{W}=\tilde{K}\int \left [ \prod_{\{e,(v)\}\in E^{\textrm{ext}}}d\tilde{\alpha}_{e} \tilde{\alpha}_{e}^{\frac{d}{2}-\Delta_{e}-1}z_{v}^{\Delta_{e}} \right ]\left [\prod_{v\in V^{\textrm{int}}} dz_{v}z_{v}^{-d-1}\right ] \frac{\tilde{\mathcal{Y}}^{R-d/2}}{\tilde{\mathcal{W}}^{R}}  \\
&\times \left [ \prod_{\{e,(v,v')\}\in E^{\textrm{int}}} d\tilde{\beta_{e}}d\tilde{\kappa_{e}} \tilde{\kappa}_{e}^{d-1-2\Delta_{e}}\tilde{\beta}_{e}^{\frac{-1}{2}}(\tilde{\kappa}_{e}-\tilde{\beta}_{e})^{\Delta_{e}-\frac{d+1}{2}}z_{v}^{\Delta_{e}}z_{v'}^{\Delta_{e}} \right ] \tilde{\mathcal{Z}}^{\frac{d|V^{\textrm{int}}|}{2}-\Delta^{\textrm{int}}-\frac{\Delta^{\textrm{ext}}}{2}}\\
&\times \delta(1 -\sum_{e} (\tilde{\beta}_{e}+\tilde{\kappa}_{e}) - \sum_{e} \tilde{\alpha}_{e} )\delta(1 -\sum_{v} z_{v})
\end{split}
\end{equation}
where 
\begin{equation}\nonumber
\begin{split}
R&=\frac{d(|E|-|V^{\textrm{int}}|)-\Delta^{\textrm{ext}}}{2} \ , \\
\tilde{K}&=\frac{\pi^{dL/2}}{2^{\Delta^{\textrm{ext}}+2\Delta^{\textrm{int}}-d|V^{\textrm{int}}|-1}}\left ( \prod_{e\in E^{\textrm{ext}}}\tilde{N}^{\textrm{ext}}_{\Delta_{e}}\right ) \left ( \prod_{e\in E^{\textrm{int}}} \tilde{N}^{\textrm{int}}_{\Delta_{e}}\right )\Gamma\left [\frac{\Delta^{\textrm{ext}}}{2}+\Delta^{\textrm{int}} -\frac{d|V^{\textrm{int}}|}{2}\right ]\Gamma \left[R\right ] \ ,
\end{split}
\end{equation}
$|E|$ is the total number of external and internal edges and $L$ is the loop order. The result looks remarkably similar to the Feynman parameterization of an off-shell Feynman diagram using Eq. (\ref{momentumpoly}), except for the additional integrals over $\tilde{\alpha}_{e}$, $\tilde{\kappa}_{e}$, and $z_{v}$.

\subsection{Polynomials from forests}\label{polynomialsfromforests}

In this section, we give an explicit formula for the $\tilde{\mathcal{W}}$ and $\tilde{\mathcal{Y}}$ polynomials for some specific Witten diagrams. We first review the combinatorial formulas for Symanzik polynomials $\mathcal{U}$ and $\mathcal{F}$, see Refs \cite{Weinzierl:2022eaz} for a more extensive review. $\mathcal{U}$ is defined as a sum over spanning trees, $T$, of the internal edges
\begin{equation}
\mathcal{U}(\tilde{\beta}_{e})=\sum_{T} \prod_{e \notin T} \tilde{\beta}_{e}
\end{equation}
where the product is over internal edges not in the spanning tree $T$. In contrast, $\mathcal{F}$ is given by a sum of $2$-forests over the internal vertices, where a $2$-forest is a forest with two connected components, 
\begin{equation}
\mathcal{F}(k_i,\tilde{\beta}_{e})=\sum_{(T_{1},T_{2})}\left (\prod_{e\notin (T_{1},T_{2})} \tilde{\beta}_{e}\right ) \left (\sum_{k_{j}\in P_{T_{1}},k_{i}\in P_{T_{2}}}k_{j}\cdot k_{i}  \right )
\end{equation}
In the above formula, $T_{1}$ and $T_{2}$ are the connected components of the $2$-forest. $P_{T}$ is the set of external momenta attached to the forest $T$.

As an illustrative example, again consider the four-point s-channel graph in Eq. (\ref{fourpoint}). The graph is tree-level, so there is no loop integration variable. There is only a single spanning tree of the internal edge, which covers the entire graph, so 
\begin{equation}
\pgfmathsetmacro{\r}{1}
\begin{tikzpicture}[baseline={([yshift=-.5ex]current bounding box.center)},every node/.style={font=\scriptsize}]
\draw [] (0,0) circle (\r cm);
\filldraw (45:\r) circle (1pt) node[above=0pt]{$3$};
\filldraw (135:\r) circle (1pt) node[above=0pt]{$2$};
\filldraw (225:\r) circle (1pt) node[below=0pt]{$1$};
\filldraw (-45:\r) circle (1pt) node[below=0pt]{$4$};
\filldraw[blue] (0:\r/2) circle (1pt) (180:\r/2) circle (1pt);
\draw [gray!30] (45:\r) -- (0:\r/2) -- (-45:\r);
\draw [gray!30] (135:\r) -- (180:\r/2) -- (225:\r);
\draw [blue,thick] (0:\r/2) -- (180:\r/2) ;
\end{tikzpicture}  \quad \leftrightarrow \quad \mathcal{U}_{s}=1 \ .
\end{equation}
In addition, there is a single $2$-forest which corresponds to two disconnected vertices, which implies 
\begin{equation}
\pgfmathsetmacro{\r}{1}
\begin{tikzpicture}[baseline={([yshift=-.5ex]current bounding box.center)},every node/.style={font=\scriptsize}]
\draw [] (0,0) circle (\r cm);
\filldraw (45:\r) circle (1pt) node[above=0pt]{$3$};
\filldraw (135:\r) circle (1pt) node[above=0pt]{$2$};
\filldraw (225:\r) circle (1pt) node[below=0pt]{$1$};
\filldraw (-45:\r) circle (1pt) node[below=0pt]{$4$};
\draw [gray!30] (45:\r) -- (0:\r/2) -- (-45:\r);
\draw [gray!30] (135:\r) -- (180:\r/2) -- (225:\r);
\draw [gray!30] (0:\r/2) -- (180:\r/2) ;
\filldraw[blue] (0:\r/2) circle (2pt) (180:\r/2) circle (2pt);
\end{tikzpicture}  \quad \leftrightarrow \quad \mathcal{F}_{s}(\tilde{\beta}_{e})=\tilde{\beta}_{1} (k_{1}+k_{2})^{2} \ .
\end{equation}
We enlarge the nodes to emphasize that each bulk vertex should be interpreted as a connected component of the $2$-forest. The result is 
\begin{equation}
\tilde{\mathcal{Y}}_{s}=1, \quad \tilde{\mathcal{W}}_{s}= \tilde{\alpha}_{1}k_{1}^{2}+\tilde{\alpha}_{2}k_{2}^{2}+\tilde{\alpha}_{3}k_{3}^{2}+\tilde{\alpha}_{4}k_{4}^{2}+\tilde{\beta}_{1}(k_{1}+k_{2})^{2} \ . 
\end{equation}
which is trivially the correct result! 

We now turn to a more non-trivial example. Consider the four-point box:
\begin{equation}
\pgfmathsetmacro{\r}{2}
\begin{tikzpicture}[baseline={([yshift=-.5ex]current bounding box.center)},every node/.style={font=\scriptsize}]
\draw [] (0,0) circle (\r cm);
\filldraw (45:\r) circle (1pt) node[above=0pt]{$3$};
\filldraw (135:\r) circle (1pt) node[above=0pt]{$2$};
\filldraw (225:\r) circle (1pt) node[below=0pt]{$1$};
\filldraw (-45:\r) circle (1pt) node[below=0pt]{$4$};
\filldraw (45:\r/2) circle (1pt) node[above=0pt]{};
\filldraw (135:\r/2) circle (1pt) node[above=0pt]{};
\filldraw (225:\r/2) circle (1pt) node[below=0pt]{};
\filldraw (-45:\r/2) circle (1pt) node[below=0pt]{};
\draw [thick] (-45:\r/2) -- (225:\r/2) -- (135:\r/2) -- (45:\r/2) -- (-45:\r/2);
\draw [thick] (-45:\r/2) -- (-45:\r);
\draw [thick] (45:\r/2) -- (45:\r);
\draw [thick] (135:\r/2) -- (135:\r);
\draw [thick] (225:\r/2) -- (225:\r);
\filldraw (0:0.45*\r) circle (0pt) node{$\tilde{\beta}_{3}$};
\filldraw (90:0.45*\r) circle (0pt) node{$\tilde{\beta}_{2}$};
\filldraw (180:0.45*\r) circle (0pt) node{$\tilde{\beta}_{1}$};
\filldraw (270:0.45*\r) circle (0pt) node{$\tilde{\beta}_{4}$};
\filldraw (45:0.85*\r) circle (0pt) node[left=0pt]{$\tilde{\alpha}_{3}$};
\filldraw (135:0.85*\r) circle (0pt) node[right=0pt]{$\tilde{\alpha}_{2}$};
\filldraw (-45:0.85*\r) circle (0pt) node[left=0pt]{$\tilde{\alpha}_{4}$};
\filldraw (-135:0.85*\r) circle (0pt) node[right=0pt]{$\tilde{\alpha}_{1}$};
\end{tikzpicture} 
\end{equation}
The four relevant spanning trees for computing $\mathcal{U}$ are given by
\begin{equation}
\pgfmathsetmacro{\r}{1}
\begin{tikzpicture}[baseline={([yshift=-.5ex]current bounding box.center)},every node/.style={font=\scriptsize}]
\draw [] (0,0) circle (\r cm);
\filldraw (45:\r) circle (1pt) node[above=0pt]{$3$};
\filldraw (135:\r) circle (1pt) node[above=0pt]{$2$};
\filldraw (225:\r) circle (1pt) node[below=0pt]{$1$};
\filldraw (-45:\r) circle (1pt) node[below=0pt]{$4$};
\filldraw (45:\r/2) circle (1pt) node[above=0pt]{};
\filldraw (135:\r/2) circle (1pt) node[above=0pt]{};
\filldraw (225:\r/2) circle (1pt) node[below=0pt]{};
\filldraw (-45:\r/2) circle (1pt) node[below=0pt]{};
\draw [gray!30] (-45:\r/2) -- (45:\r/2);
\draw [gray!30] (-45:\r/2) -- (-45:\r);
\draw [gray!30] (45:\r/2) -- (45:\r);
\draw [gray!30] (135:\r/2) -- (135:\r);
\draw [gray!30] (225:\r/2) -- (225:\r);
\draw [blue,thick] (-45:\r/2) -- (225:\r/2) -- (135:\r/2) -- (45:\r/2);
\end{tikzpicture} \quad \leftrightarrow \quad \mathcal{U}\ni \tilde{\beta}_{3} \ , 
\end{equation}
and rotations thereof. The end result is  
\begin{equation}
\mathcal{U}_{\textrm{box}}=\tilde{\beta}_{1}+\tilde{\beta}_{2}+\tilde{\beta}_{3}+\tilde{\beta}_{4} \ .
\end{equation}
There are two topologies of $2$-forests that contribute to $\mathcal{F}$:
\begin{equation}
\pgfmathsetmacro{\r}{1}
\begin{tikzpicture}[baseline={([yshift=-.5ex]current bounding box.center)},every node/.style={font=\scriptsize}]
\draw [] (0,0) circle (\r cm);
\filldraw (45:\r) circle (1pt) node[above=0pt]{$3$};
\filldraw (135:\r) circle (1pt) node[above=0pt]{$2$};
\filldraw (225:\r) circle (1pt) node[below=0pt]{$1$};
\filldraw (-45:\r) circle (1pt) node[below=0pt]{$4$};
\filldraw (45:\r/2) circle (1pt) node[above=0pt]{};
\filldraw (135:\r/2) circle (1pt) node[above=0pt]{};
\filldraw (225:\r/2) circle (1pt) node[below=0pt]{};
\filldraw (-45:\r/2) circle (1pt) node[below=0pt]{};
\draw [blue,thick] (-45:\r/2) -- (225:\r/2);
\draw [gray!30] (225:\r/2) -- (135:\r/2);
\draw [blue,thick] (135:\r/2) -- (45:\r/2);
\draw [gray!30] (-45:\r/2) -- (45:\r/2);
\draw [gray!30] (-45:\r/2) -- (-45:\r);
\draw [gray!30] (45:\r/2) -- (45:\r);
\draw [gray!30] (135:\r/2) -- (135:\r);
\draw [gray!30] (225:\r/2) -- (225:\r);
\filldraw[blue] (45:\r/2) circle (2pt) (135:\r/2) circle (2pt);
\filldraw[blue] (-45:\r/2) circle (2pt) (-135:\r/2) circle (2pt);
\end{tikzpicture} \quad \leftrightarrow \quad \mathcal{F}\ni \tilde{\beta}_{1}\tilde{\beta}_{3}(k_{2}+k_{3})^{2}
\end{equation}
\begin{equation}\label{forestex}
\pgfmathsetmacro{\r}{1}
\begin{tikzpicture}[baseline={([yshift=-.5ex]current bounding box.center)},every node/.style={font=\scriptsize}]
\draw [] (0,0) circle (\r cm);
\filldraw (45:\r) circle (1pt) node[above=0pt]{$3$};
\filldraw (135:\r) circle (1pt) node[above=0pt]{$2$};
\filldraw (225:\r) circle (1pt) node[below=0pt]{$1$};
\filldraw (-45:\r) circle (1pt) node[below=0pt]{$4$};
\filldraw (45:\r/2) circle (1pt) node[above=0pt]{};
\filldraw (135:\r/2) circle (1pt) node[above=0pt]{};
\filldraw (225:\r/2) circle (1pt) node[below=0pt]{};
\filldraw (-45:\r/2) circle (1pt) node[below=0pt]{};
\draw [blue,thick] (-45:\r/2) -- (225:\r/2);
\draw [blue,thick] (225:\r/2) -- (135:\r/2);
\draw [gray!30] (135:\r/2) -- (45:\r/2);
\draw [gray!30] (-45:\r/2) -- (45:\r/2);
\draw [gray!30] (-45:\r/2) -- (-45:\r);
\draw [gray!30] (45:\r/2) -- (45:\r);
\draw [gray!30] (135:\r/2) -- (135:\r);
\draw [gray!30] (225:\r/2) -- (225:\r);
\filldraw[blue] (45:\r/2) circle (2pt) (135:\r/2) circle (2pt);
\filldraw[blue] (-45:\r/2) circle (2pt) (-135:\r/2) circle (2pt);
\end{tikzpicture} \quad \leftrightarrow \quad \mathcal{F}\ni\tilde{\beta}_{1}\tilde{\beta}_{4} k_{3}^{2} \ . 
\end{equation}
We emphasize that the isolated dot should be regarded as the second connected component of the tree in Eq.~(\ref{forestex}). The other contributing $2$-forests can be derived by cyclic permutations of the above, resulting in 
\begin{equation}
\begin{split}
\mathcal{F}_{\textrm{box}}&=\tilde{\beta}_{1}\tilde{\beta}_{3}(k_{2}+k_{3})^{2}+\tilde{\beta}_{2}\tilde{\beta}_{4}(k_{3}+k_{4})^{2}\\
&+\tilde{\beta}_{1}\tilde{\beta}_{4} k_{3}^{2} +\tilde{\beta}_{2}\tilde{\beta}_{1} k_{4}^{2}+\tilde{\beta}_{2}\tilde{\beta}_{1} k_{4}^{2}+\tilde{\beta}_{3}\tilde{\beta}_{2} k_{1}^{2}\ .
\end{split}
\end{equation}
Combining these formulas for $\mathcal{U}$ and $\mathcal{F}$, we can compute $\tilde{\mathcal{Y}}$ and $\tilde{\mathcal{W}}$:
\begin{equation}
\begin{split}
\tilde{\mathcal{Y}}_{\textrm{box}}&=\tilde{\beta}_{1}+\tilde{\beta}_{2}+\tilde{\beta}_{3}+\tilde{\beta}_{4} \ . \\
\tilde{\mathcal{W}}_{\textrm{box}}&=\left (\tilde{\beta}_{1}+\tilde{\beta}_{2}+\tilde{\beta}_{3}+\tilde{\beta}_{4}\right )\left (\tilde{\alpha}_{1}k_{1}^{2}+\tilde{\alpha}_{2}k_{2}^{2}+\tilde{\alpha}_{3}k_{3}^{2}+\tilde{\alpha}_{4}k_{3}^{2}\right ) \\
&+\tilde{\beta}_{1}\tilde{\beta}_{3}(k_{2}+k_{3})^{2}+\tilde{\beta}_{2}\tilde{\beta}_{4}(k_{3}+k_{4})^{2} \\
&+\tilde{\beta}_{1}\tilde{\beta}_{4} k_{3}^{2} +\tilde{\beta}_{2}\tilde{\beta}_{1} k_{4}^{2}+\tilde{\beta}_{2}\tilde{\beta}_{1} k_{4}^{2}+\tilde{\beta}_{3}\tilde{\beta}_{2} k_{1}^{2} \ .
\end{split}
\end{equation}

\section{Application: generalization of Weinberg's theorem on UV convergence}\label{wittenrevmomUV}

We can leverage the momentum-space WF parameterization to prove a generalization of Weinberg's theorem on UV convergence of Feynman integrals \cite{Weinberg:1959nj}. Weinberg's theorem on UV convergence states that a Feynman diagram is UV convergent if 
\begin{equation}\label{degreeofdiv}
D_{g}=|E|_{g}-\frac{(d+1)L_{g}}{2}>0
\end{equation}
for all subgraphs $g$, where $|E|_{g}$ is the number of edges in the subgraph and $L_{g}$ is the loop order of the sub-graph.\footnote{Remember that the spacetime dimension is $d+1$ in our conventions, not $d$.} Weinberg's theorem on UV convergence of Feynman integrals was crucial for proving the validity of renormalization in perturbation theory. Witten diagrams are, of course, expected to exhibit the same UV divergences as Feynman integrals. The short-distance behavior of the theory governs UV divergences, making them insensitive to the large-scale structure of spacetime. Nevertheless, it is satisfying that the WF parameterization makes this intuition precise in a quantitative way.

The goal is to investigate whether contributions from specific regions of the integration domain exhibit divergent behavior. The integrand simplifies in these integration domains. As a toy example, consider the integral,
\begin{equation}
F(A,B)=\int_{0}^{\infty}dx x^{A-1}P^{B}, \quad P=(1+z_{1}x+z_{2}x^{2}) \ .
\end{equation}
We want to know when the above integral is divergent. There can be a divergence in the $x\rightarrow 0$ region of the integration domain. In this region, we can approximate the polynomial as $P \rightarrow 1$, and the integral becomes
\begin{equation}
F(A,B) \supset \int_{0}^{\Lambda} dx x^{A-1} = \frac{\Lambda^{A-1}}{A}
\end{equation}
and is divergent if $A\leq 0$. Importantly, we did not need the full behavior of the $P$ integral to analyze this divergence; only the dominant monomial in the $x\rightarrow 0$ limit was required. We can also consider the $x\rightarrow \infty $ limit, where $P\rightarrow z_{2}x^{2}$. The relevant integral is 
\begin{equation}
F(A,B) \supset \int^{\infty}_{\Lambda} dx x^{A-1+2B}=\frac{-\log(\Lambda)}{A+2B} \ , 
\end{equation}
which converges only when $2B < -A$. Therefore, we can check convergence by studying the behavior of the integrand in specific limits. Approximating polynomials by the dominant monomial in the sum, known as the \emph{tropical approximation}, is discussed in Ref.~\cite{Arkani-Hamed:2022cqe} in the context of Feynman integrals.\footnote{See Sec.~4.1 of Ref.~\cite{Herderschee:2021dez} for a brief introduction to the tropical approximation of polynomials.}

We will study the contribution from the integration domain where the $\tilde{\beta}_{e}$ of a particular sub-graph, $g$, are very small so that we can approximate the polynomials $\mathcal{W}$ and $\mathcal{Y}$ by the dominant terms in the sum. We will additionally assume that the $z_{v}$ of the subgraph are localized,
\begin{equation}
\forall \{e,(v,v') \} \in g : \quad (z_{v}-z_{v'})^{2}\sim \tilde{\beta}_{e}\ ,
\end{equation}
so that the $(z_{v}-z_{v'})^{2}\tilde{\beta}_{e}^{-1}$ terms in $\tilde{\mathcal{Z}}$ are order one. To be more explicit, we make the change of variables 
\begin{equation}
z_{v}\rightarrow z_{v'}+\sqrt{\tilde{\beta}_{e}}\delta z_{v'}
\end{equation}
which leads to a Jacobian factor that is a monomial in the $\tilde{\beta}_{e}$ with $e\in g$ whose total degree is $(|E_{g}|-L_{g})/2$; that is 
\begin{equation}
\textrm{Jacobian}\propto \prod_{e\in g} \tilde{\beta}_{e}^{n_{e}}, \quad \textrm{with} \quad \sum_{e\in g}n_{e}=\frac{|E_{g}|-L_{g}}{2} \ .
\end{equation}
For example, consider the sub-polynomial in $\mathcal{Z}$ associated with a triangle sub-graph:
\begin{equation}\label{trianglesubgrah}
\mathcal{Z}\supset \frac{(z_{1}-z_{2})^{2}}{\tilde{\beta}_{1}}+ \frac{(z_{2}-z_{3})^{2}}{\tilde{\beta}_{2}}+\frac{(z_{1}-z_{3})^{2}}{\tilde{\beta}_{3}} \ . 
\end{equation}
We are taking $\tilde{\beta}_{1},\tilde{\beta}_{2},\tilde{\beta}_{3}$ to be small. Therefore, we make the change of variables
\begin{equation}
z_{2}=z_{1}+\sqrt{\tilde{\beta}_{1}}\delta z_{2}, \quad z_{3}=z_{1}+\sqrt{\tilde{\beta}_{3}}\delta z_{3}
\end{equation}
which leads to a Jacobian factor of $\sqrt{\tilde{\beta_{1}}\tilde{\beta}_{3}}$. Upon taking the $\tilde{\beta}_{e}$ small, one finds that Eq. (\ref{trianglesubgrah}) becomes
\begin{equation}
\mathcal{Z}\supset (\delta z_{2})^{2}+ (\delta z_{3})^{2}+\frac{(\sqrt{\tilde{\beta}_{1}}\delta z_{2}-\sqrt{\tilde{\beta}_{3}}\delta z_{3})^{2}}{\tilde{\beta}_{2}}
\end{equation}
which is $\mathcal{O}(1)$ and can be safely ignored. Finally, we assume that all other variables are $\mathcal{O}(1)$. We will explicitly check whether the contribution to the integral from this region is divergent. We note that our assumptions mean this analysis will miss divergences associated with taking the $\tilde{\kappa}$, $\tilde{\alpha}$, and/or $z_{v}$ variables small or large.

The trick here is that the WF parameterized integrand exhibits the same behavior as in flat space because the $\mathcal{U}$ and $\mathcal{F}$ polynomials are the same. We consider the AdS integrals here, but note the same argument applies to dS. It is convenient to shift to a Lee-Pomeransky-like representation:
\begin{equation}\label{LPmomentum}
\begin{split}
&\tilde{W}\propto \int \left [ \prod_{\{e,(v)\}\in E^{\textrm{ext}}}d\tilde{\alpha}_{e} \tilde{\alpha}_{e}^{\frac{d}{2}-\Delta_{e}-1}z_{v}^{\Delta_{e}} \right ]\left [\prod_{v\in V^{\textrm{int}}} dz_{v}z_{v}^{-d-1}\right ] (\tilde{\mathcal{Y}}+\tilde{\mathcal{W}})^{-d/2} \\
&\times \left [ \prod_{\{e,(v,v')\}\in E^{\textrm{int}}} d\tilde{\beta_{e}}d\tilde{\kappa_{e}} \tilde{\kappa}_{e}^{d-1-2\Delta_{e}}\tilde{\beta}_{e}^{\frac{-1}{2}}(\tilde{\kappa}_{e}-\tilde{\beta}_{e})^{\Delta_{e}-\frac{d+1}{2}}z_{v}^{\Delta_{e}}z_{v'}^{\Delta_{e}} \right ] \tilde{\mathcal{Z}}^{\frac{d|V^{\textrm{int}}|}{2}-\Delta^{\textrm{int}}-\frac{\Delta^{\textrm{ext}}}{2}} \\
&\times \delta(1-\sum_{v}z_{v}) \ ,
\end{split}
\end{equation}
where we are ignoring the numerical pre-factor because it is unimportant. The $\tilde{\kappa}_{e}$ and $\tilde{\alpha}_{e}$ integration variables are integrated from zero to infinity in the above expression. In contrast, $\tilde{\beta}_{e}$ is integrated from zero to $\tilde{\kappa}_{e}$. To show this representation is equivalent to the WF parameterization, insert the delta-function 
\begin{equation}
\forall \tilde{\alpha}_{e},\tilde{\beta}_{e},\tilde{\kappa}_{e}>0:\quad \int d\lambda (1-\sum \tilde{\alpha}_{e}-\sum (\tilde{\beta}_{e}+\tilde{\kappa}_{e}))
\end{equation}
into Eq. (\ref{LPmomentum}), re-scale all the $\tilde{\alpha}_{e},\tilde{\beta}_{e},\tilde{\kappa}_{e}\rightarrow \lambda \tilde{\alpha}_{e},\lambda \tilde{\beta}_{e},\lambda \tilde{\kappa}_{e}$, re-scale $\lambda \rightarrow (\tilde{\mathcal{Y}}/\tilde{\mathcal{W}})\lambda$ and then integrate over $\lambda$. 

To demonstrate that the above integral shares the same UV behavior as the corresponding flat-space integral, we rewrite the polynomial as
\begin{equation}\label{polynomia}
\tilde{\mathcal{Y}}+\tilde{\mathcal{W}}=\mathcal{U}(\tilde{\beta}_{e})\left(1+ \mathcal{S}(k_{i}^{2},\tilde{\alpha}_{e})\right )+\mathcal{F}(k_i,\tilde{\beta}_{e}) \ .
\end{equation}
The polynomial in Eq. (\ref{polynomia}) has the same behavior in the given $\tilde{\beta}_{e}\rightarrow 0$ limit as the Lee-Pomeransky representation of the corresponding flat space Feynman integral. Furthermore, remember that taking the $z_{v}$ of the sub-graph to be highly localized leads to additional factors of $\tilde{\beta}_{e}$, leading to the result
\begin{equation}\label{divintegral}
\begin{split}
\tilde{W} \propto & (\ldots) \int \left [ \prod_{e\in g}d\tilde{\beta}_{e}\tilde{\beta}_{e}^{-\frac{1}{2}+n_{e}}\right ] \lim_{\forall e\in g; \ \tilde{\beta}_{e}\rightarrow 0} (\mathcal{U}(\tilde{\beta}_{E})\left(1+ \mathcal{S}(k_{i}^{2},\tilde{\alpha}_{e})\right )+\mathcal{F}(k_i,\tilde{\beta}_{e}))^{-d/2} \ , \\
&\quad \textrm{with}\quad \sum_{e\in g}n_{e}=\frac{|E_{g}|-L_{g}}{2} \ .
\end{split}
\end{equation}
The $(\ldots)$ contains an integral over the other parameters and is $\mathcal{O}(1)$ by assumption. Using the behavior of the Lee-Pomeransky polynomial when $\tilde{\beta}_{e}$ go to zero, see Ref. \cite{Arkani-Hamed:2022cqe}, the integral in Eq. (\ref{divintegral}) is divergent whenever 
\begin{equation}\label{UVfincond}
D_{g}=|E|_{g}-\frac{(d+1)L_{g}}{2}\leq 0
\end{equation}
which is the same as Eq. (\ref{degreeofdiv}). We have therefore proven a generalization of Weinberg's theorem \cite{Weinberg:1959nj} in AdS. Suppose all sub-diagrams, $g$, in the Witten diagram obey $D_{g}>0$. The Witten diagram is then convergent in the small $\tilde{\beta}$ region of integration, which we refer to as the UV region. Note that we have not proven the full integral is convergent. We have only proven it is free of a specific type of UV divergence. For example, a given Feynman integral in flat space can diverge in the infrared (IR) even if it is UV finite.

It is also instructive to comment on IR divergences. In flat space, amplitudes can develop IR divergences from regions where the loop momentum becomes very small. Within the Feynman parameterization, these divergences correspond to regions of integration space where the variables $\tilde{\beta}_{e}$ grow large. By contrast, AdS correlators are free of such long-distance divergences, since AdS effectively acts as a confining box \cite{Fitzpatrick:2011jn}. In the Witten-Feynman parameterization this feature is manifest. The integrals over $\tilde{\beta}_{e}$ are bounded from above by $\tilde{\kappa}_{e}$, making explicit how AdS regulates the IR behavior that would otherwise diverge in flat space.

The above arguments apply only to AdS. In dS, the reasoning is effectively the same, with slight modifications to the contour and exponents. One still obtains Eq.~(\ref{divintegral}) and, subsequently, the condition in Eq.~(\ref{UVfincond}), for the Witten diagrams relevant for in-in correlators. In contrast to AdS diagrams, when computing the dS in-in correlators, the $\tilde{\beta}_{e}$ are NOT bounded from above, which reflects how dS correlation functions can exhibit IR divergences reminiscent of those in flat space.

\section{Application: series representations}\label{diffeqAdS}

As a final application, we note that we can algorithmically derive series representations for the Witten diagrams of (A)dS. We will first derive the series representation of a simple hypergeometric function to illustrate the general computation strategy. We will then apply the result to the WF parameterized three-point Witten relevant for BFSS, reproducing results in Ref. \cite{Bobev:2025idz}. Finally, we discuss the application of this series representation to the four-point s-channel diagram, but we do not evaluate the sum. 

\subsection{Warm-up}\label{warmupgkz}

As an introductory example, consider the function 
\begin{equation}\label{originalexample}
F(z_{i})=\int_{0}^{\infty} dx x^{-r_{1}-1} (z_{1}+z_{2}x+z_{3}x^{2})^{r_{0}} \ .
\end{equation}
The first step is using the Mellin-Barnes transformation,
\begin{equation}
(z_{1}+z_{2}x+z_{3}x^{2})^{r_{0}}=\frac{(2\pi i)^{-2}}{\Gamma(-r_{0})}\int \frac{d\sigma_{1}d\sigma_{2}d\sigma_{3} \ z_{1}^{\sigma_{1}}z_{2}^{\sigma_{2}}z_{3}^{\sigma_{3}}}{[\Gamma(-\sigma_{1})\Gamma(-\sigma_{2})\Gamma(-\sigma_{3})]^{-1}}  x^{\sigma_{2}+2\sigma_{3}}\delta(\sigma_{1}+\sigma_{2}+\sigma_{3}-r_{0})\ ,
\end{equation}
to replace the polynomial with a collection of integrals over $\sigma_{i}$. We then evaluate the integral over $x$ using the identity:
\begin{equation}
\int_{0}^{\infty} \frac{dx}{x} x^{A}\equiv 2\pi i \delta(A) \ .
\end{equation}
The end result is 
\begin{equation}\label{Mellinbarnesform}
F(z_{i})=\frac{(2\pi i)^{-1}}{\Gamma(-r_{0})} \int \frac{d\sigma_{1}d\sigma_{2}d\sigma_{3}}{[\Gamma(-\sigma_{1})\Gamma(-\sigma_{2})\Gamma(-\sigma_{3})]^{-1}}z_{1}^{\sigma_{1}}z_{2}^{\sigma_{2}}z_{3}^{\sigma_{3}} \delta(\sigma_{1}+\sigma_{2}+\sigma_{3}-r_{0})\delta(2\sigma_{3}+\sigma_{1}-r_{1}) \ .
\end{equation}
We evaluate the delta functions assuming that $r$ and $\sigma_{i}$ are real. After performing this evaluation, we can use the standard Mellin-Barnes contour, reducing the integral to a sum over residues.

We write the delta functions explicitly to provide an explicit formula for the residues that contribute to the sum. The residues of the $\sigma$ correspond to vectors satisfying the delta function constraints in Eq.~(\ref{Mellinbarnesform}). The delta-function constraints impose the condition:
\begin{equation}\label{matrixsum}
\begin{pmatrix}
r_{0} \\
r_{1}
\end{pmatrix}=\begin{pmatrix}
1 & 1 & 1 \\
0 & 1 & 2 \\
\end{pmatrix}\cdot \begin{pmatrix}
n_{1} \\
n_{2} \\
n_{3}
\end{pmatrix} \ .
\end{equation}
One way to express the sum over residues is 
\begin{equation}
F(z_{i})=\sum_{n\in \mathbb{Z}} \textrm{Res}_{n=n'}\left [ \prod_{j=1}^{3}\frac{z_{j}^{n_{j}'}}{\Gamma(n_{j}'+1)} \right ]
\end{equation}
where the sum is over solutions to Eq. (\ref{matrixsum}) and the residue is taken on the support of the delta-functions. We can express the space of solutions to Eq. (\ref{matrixsum}) in terms of a sum over the null vectors of $\mathcal{A}$
\begin{equation}\label{matrixsum2}
0=\begin{pmatrix}
1 & 1 & 1 \\
0 & 1 & 2 \\
\end{pmatrix}\cdot \begin{pmatrix}
\gamma_{1} \\
\gamma_{2} \\
\gamma_{3}
\end{pmatrix} \ .
\end{equation}
Therefore, the sum is really over particular families of null vectors of the matrix in Eq. (\ref{matrixsum2}). 

Let us consider an explicit evaluation of the function when $r_{1}=-1$ and $r_{0}=-2$. We parameterize the sum over $n$ as 
\begin{equation}
\forall n_{2}\in\mathbb{Z}_{\geq 0}: \quad n=(-\frac{3+n_{2}}{2},n_{2},-\frac{1+n_{2}}{2})
\end{equation}
so the sum becomes 
\begin{equation}
F(z_{i})=\frac{1}{z_{1}^{3/2}z_{3}^{1/2}}\sum_{\sigma_{2}\geq 0}\frac{\sqrt{\pi } \left(-\frac{1}{2}\right)^{\sigma _2} \Gamma \left(\frac{1}{2} \left(\sigma _2+3\right)\right) }{\sigma _2 \Gamma \left(\frac{\sigma _2}{2}\right)} \left (  \frac{z_{2}^{2}}{z_{1}z_{3}}\right)^{\sigma_{2}/2}
\end{equation}
which is convergent when $|z_{2}/\sqrt{z_{1}z_{3}}|$ is sufficiently small. Note that this is not the only family of residues we could have chosen. We could have also chosen 
\begin{equation}
\quad n_{3}\in \mathbb{Z}_{\geq 0}:\quad n=(-1+\sigma_{3},-1-2\sigma_{3},\sigma_{3}) \ ,
\end{equation}
which yields the sum 
\begin{equation}
\begin{split}
F(z_{i})&=\frac{1}{z_{1}z_{2}}\sum_{\sigma_{3}\geq 0}\frac{4^{\sigma _3} \Gamma \left(\sigma _3+\frac{1}{2}\right) }{\sqrt{\pi } \Gamma \left(\sigma _3\right)}  \left (  \frac{z_{1}z_{3}}{z_{2}}\right)^{\sigma_{3}}\\
&\times \left(2 H_{2 \sigma _3}-\frac{1}{\sigma _3}-2 \left(\psi ^{(0)}\left(\sigma _3\right)+\gamma \right)+\log \left(\frac{z_1 z_3}{z_2^2}\right)\right) \ , 
\end{split}
\end{equation}
which converges when $z_{2}/\sqrt{z_{1}z_{3}}$ is sufficiently large.  

This strategy applies to any generalized Euler integral. For a generalized Euler integral, the matrix in Eq. (\ref{matrixsum}) is called the $\mathcal{A}$-matrix of the system \cite{gel1989hypergeometric,gelfand1993correction}. We compute it by identifying the columns of the matrix with terms in the polynomials entering the integral.
The columns correspond to exponent vectors of the integration variables. The benefit of the above computation strategy is that, for a generic choice of $z_{i}$, we expect there should be a choice of convergent series representation.

\subsection{Example: three-point correlator in BFSS}\label{bfss3p}

In this section, we study the tree-level three-point correlator of the BFSS matrix model. Correlators in the BFSS matrix model lack conformal symmetry, so the kinematic dependence of the three-point function is not fixed and instead yields a nontrivial function.

Although Refs.~\cite{Biggs:2025qfh,Bobev:2025idz} recently computed this correlator, we include it here to illustrate the utility of the WF parameterization.
The WF parameterization of the relevant Witten diagram yields
\begin{equation}
\begin{split}
& \begin{tikzpicture}[baseline={([yshift=-.5ex]current bounding box.center)},every node/.style={font=\scriptsize}]\pgfmathsetmacro{\r}{0.8}
\draw [] (0,0) circle (\r cm);
\tikzset{decoration={snake,amplitude=.4mm,segment length=1.5mm,post length=0mm,pre length=0mm}}
\filldraw (0:\r) circle (1pt) node[right=0pt]{$1$};
\filldraw (120:\r) circle (1pt) node[above=0pt]{$2$};
\filldraw (240:\r) circle (1pt) node[below=0pt]{$3$};
\draw [thick] (0:\r) -- (0:0);
\draw [thick] (120:\r) -- (120:0);
\draw [thick] (240:\r) -- (240:0);
\end{tikzpicture} \propto \int [\prod_{i=1}d\alpha_{i}\alpha _i^{\Delta _i-\frac{19}{10}}] \frac{\mathcal{W}^{r_{0}}}{\mathcal{Y}^{r_{1}}}\delta(1-\mathbf{m}_{1}\alpha_{1}-\mathbf{m}_{2}\alpha_{2}-\mathbf{m}_{3}\alpha_{3}) \ ,
\end{split}
\end{equation}
where 
\begin{equation}
\begin{split}
r_{1}&=\frac{9}{10}\\
r_{0}&=\frac{9}{5}-\frac{\Delta^{\textrm{ext}}}{2} \\
\mathcal{Y}&=\alpha_{1}+\alpha_{2}+\alpha_{3} \\
\mathcal{W}&=x_{12}^{2}\alpha_{1}\alpha_{2}+x_{13}^{2}\alpha_{1}\alpha_{3}+x_{23}^{2}\alpha_{2}\alpha_{3} \\
\end{split}
\end{equation}
Note that the $\mathcal{Z}$ polynomial is reduced to a factor of $\mathcal{Y}$ because there is only a single bulk vertex. Furthermore, we have left the delta function more general than the previous formula as in Eq. (\ref{gendeltafun}).

We will employ a strategy similar to the one used in the previous section. First, we choose $\vec{\mathbf{m}}=(1,0,0)$, which sets $\alpha_{1}=1$. We then apply the Mellin-Barnes transformation to the $\mathcal{W}$ polynomial, but not the $\mathcal{Y}$ polynomial. The result is
\begin{equation}
\begin{split}
W=\frac{1}{\Gamma[-r_{0}] (2\pi i)^{2}} &\int \frac{d\sigma_{1}d\sigma_{2}d\sigma_{3}}{[\Gamma(-\sigma_{1})\Gamma(-\sigma_{2})\Gamma(-\sigma_{3})]^{-1}} \int d\alpha_{2}d\alpha_{3}\frac{\alpha_{2}^{\Delta_{2}-1+\sigma_{1}+\sigma_{3}}\alpha_{3}^{\Delta_{3}-1+\sigma_{2}+\sigma_{3}}}{\mathcal{Y}^{r_{1}}}  \\
&\times  x_{12}^{\sigma_{1}}x_{13}^{\sigma_{2}} x_{23}^{\sigma_{3}}\delta(r_{0}-\sigma_{1}-\sigma_{2}-\sigma_{3}) \ .
\end{split}
\end{equation}
Fortunately, the integral over $\alpha_{2}$ and $\alpha_{3}$ is known in closed form 
\begin{equation}
\begin{split}
&\int d\alpha_{2}d\alpha_{3}\frac{\alpha_{2}^{\Delta_{2}-1+\sigma_{1}+\sigma_{3}}\alpha_{3}^{\Delta_{3}-1+\sigma_{2}+\sigma_{3}}}{(1+\alpha_{2}+\alpha_{3})^{\frac{9}{10}}} \\
&=\frac{\Gamma \left(-\Delta _2-\Delta _3-\sigma _1-\sigma _2-2 \sigma _3+\frac{9}{10}\right) \Gamma \left(\Delta _2+\sigma _1+\sigma _3\right) \Gamma \left(\Delta _3+\sigma _2+\sigma _3\right)}{\Gamma \left(\frac{9}{10}\right)}
\end{split}
\end{equation}
We then have a double sum over residues over the product of six gamma functions. Depending on the choice of residues we sum over, we obtain different series representations with distinct regions of convergence. We choose to solve for $\sigma_{3}$ and sum over the family of residues
\begin{equation}
\begin{split}
\sigma_{1}&=\frac{9}{10}-\frac{\Delta_{1}+\Delta_{2}-\Delta_{3}}{2}+n_{1}, \quad n_{1} \\
\sigma_{2}&=\frac{9}{10}-\frac{\Delta_{1}+\Delta_{3}-\Delta_{2}}{2}+n_{2}, \quad n_{2} \\
\end{split}
\end{equation}
where $n_{1},n_{2}\in \mathbb{Z}_{\geq 0}$. These four families of residues produce four separate sums.
However, all the sums are series expansions in 
\begin{equation}
\left (\frac{x_{1,2}^{2}}{x_{2,3}^{2}}\right )^{n_{2}}\left (\frac{x_{1,3}^{2}}{x_{2,3}}\right )^{n_{1}}
\end{equation}
and converge very quickly if $(x_{1,2}^{2}/x_{2,3}^{2}),(x_{1,3}^{2}/x_{2,3}^{2})$ are small. Note that it is clear from permutation symmetry that we could have also written a sum that converges when either $(x_{2,3}^{2}/x_{1,2}^{2}),(x_{1,3}^{2}/x_{1,2}^{2})$ or $(x_{2,3}^{2}/x_{1,3}^{2}),(x_{1,2}^{2}/x_{1,3}^{2})$ are small. This difference amounts to a choice of what family of residues we sum over. For a given choice of $x_{i,j}^{2}$, we should choose a particular family of residues that converges. 

We note that the individual double series appearing in the result correspond to the Appell hypergeometric function  
\begin{equation}
F_{4}(a,b;c,d;z_{1},z_{2}) = \sum_{m,n=0}^{\infty} \frac{(a)_{m+n} (b)_{m+n}}{(c)_{m} (d)_{n}} \frac{z_1^m z_2^n}{m! n!} \ ,
\end{equation}
so the full expression can be written as a sum of four $F_4$ functions, as shown in Ref.~\cite{Bobev:2019bvq}. However, the philosophy of this section is somewhat orthogonal to that approach. For more complicated Witten diagrams, the result will almost certainly not reduce to any known hypergeometric functions implemented in \textit{Mathematica}. Indeed, even the $F_4$ function itself was only introduced in \textit{Mathematica} version 13.3. The challenge of computing higher-order Witten diagrams raises natural questions: What makes $F_4$ special compared to other multivariable functions? And in what sense can we claim to have ``solved'' a function?

Here, we adopt the viewpoint that a function is ``solved'' if its numerical value can be evaluated for arbitrary inputs. From this perspective, $F_4$ is considered solved because \textit{Mathematica} can numerically evaluate it for general inputs. Similarly, we hope the series strategy developed in this section should provide a solution of this type for \textit{any} Witten diagram, in the sense that there should exist a convergent series expansion for any point in the kinematic space. However, these series may be cumbersome in practice. The number of nested summations equals the rank of $\ker \mathcal{A}$, which can grow rapidly with the complexity of the diagram.

\subsection{Example: sketch of four-point s-channel}

We can now consider the four-point s-channel graph using a similar strategy. The goal of this section is to provide computational tricks that help simplify this computation. However, we do not evaluate the sum, either numerically or analytically. 

The WF parameterization is not quite suitable for the computation strategy in Section \ref{warmupgkz} for the single reason that the $\alpha$, $\beta$, and $\kappa$ variables are not integrated from zero to infinity. Instead, we consider a Lee-Pomeransky-like representation, which is of the form 
\begin{equation}\label{LPWFrepsketchfull}
\begin{split}
&W=K'\int \left [ \prod_{e\in E^{\textrm{ext}}}d\alpha_{e} \alpha_{e}^{\Delta_{e}-1}z^{\Delta_{e}} \right ]\left [\prod_{v\in V^{\textrm{int}}} dz_{v}z_{v}^{-d-1}\right ]  \\
&\times \left [ \prod_{e\in E^{\textrm{int}}} d\beta_{e}d\kappa_{e} \beta_{e}^{d-1-\Delta_{e}}(\kappa_{e}(\beta_{e}-\kappa))^{\Delta_{e}-\frac{d+1}{2}}\prod_{v\in e}z_{v}^{\Delta_{e}} \right ] \mathcal{G}^{\frac{1}{2}(d-\Delta^{\textrm{ext}}-2\Delta^{\textrm{int}})} \ .
\end{split}
\end{equation}
where
\begin{equation}
\begin{split}
\mathcal{G}&=\mathcal{Z}_{s}+\mathcal{Y}_{s}+\mathcal{F}_{s}
\end{split}
\end{equation}
and
\begin{equation}
K'=K\times \frac{2 \Gamma \left[\Delta^{\textrm{int}}+\frac{1}{2} (\Delta^{\textrm{ext}}-d)\right]}{\Gamma \left[\frac{\Delta^{\textrm{ext}}}{2}\right] \Gamma \left[\frac{1}{2} (d-\Delta^{\textrm{ext}})\right] \Gamma \left[-d+\Delta^{\textrm{int}}+\frac{\Delta^{\textrm{ext}}}{2}\right]} \ .
\end{equation}
One can show that the above integral is equivalent to the WF parameterization by inserting the delta functions,
\begin{equation}
\begin{split}
1&=\int d\lambda_{1}\ \delta(\lambda_{1}-\sum (\beta_{e}+\kappa_{e})-\sum \alpha_{e}) \ , \\
1&=\int d\lambda_{2}\ \delta(\lambda_{2}-\sum z_{v}) \ ,
\end{split}
\end{equation}
and performing the u-sub
\begin{equation}
\alpha_{e},\beta_{e},\kappa_{e}\rightarrow \lambda_{1}\alpha_{e},\lambda_{1}\beta_{e},\lambda_{1}\kappa_{e}, \quad z_{v}\rightarrow \lambda_{2}z_{v} \ .
\end{equation}
Unlike the previous derivation, one should now perform the u-sub 
\begin{equation}
\lambda_{2}\rightarrow \frac{1}{\sqrt{\lambda_{1}\mathcal{Z}}}\lambda_{2}, \quad
\end{equation}
and integrate out $\lambda_{2}$. After $\lambda_{2}$ is gone, one performs the u-sub 
\begin{equation}
\lambda_{1}\rightarrow \frac{\mathcal{Y}}{\mathcal{W}}\lambda_{1}
\end{equation}
and integrates out $\lambda_{1}$. The result is the original position-space WF parameterization. 

Given the Lee-Pomeransky-like representation of the Witten diagram, we can now apply the computation strategy in the previous section. However, it is helpful to further simplify the integral first by leveraging conformal symmetry as described in Section \ref{levconf}. This step reduces the number of terms in $\mathcal{G}$, which in turn reduces the number of nested sums in the resulting series representation. After applying the procedure described in Section \ref{levconf}, the WF polynomials are 
\begin{align}
\mathcal{W}_{s}&\rightarrow \alpha _1 \alpha _3 \beta  \eta +\alpha _3 \alpha _2 \beta  \xi +\alpha _1 \alpha _2 \beta +\alpha _4 \alpha _2 \beta +\alpha _1 \alpha _4 \beta +\alpha _3 \alpha _4 \beta +\alpha _1 \alpha _3 \alpha _2+\alpha _3 \alpha _4 \alpha _2+\alpha _1 \alpha _3 \alpha _4\nonumber \ , \\
\mathcal{Y}_{s}&\rightarrow\alpha _1 \beta +\alpha _2 \beta +\alpha _3 \beta +\alpha _3 \alpha _1+\alpha _2 \alpha _3\ ,  \\
\mathcal{Z}_{s}&\rightarrow\alpha _1 z_1^2+\alpha _2 z_1^2+\alpha _3 z_2^2+\beta  z_1^2-2 \beta  z_2 z_1+\beta  z_2^2+4 \kappa  z_2 z_1  \ . \nonumber 
\end{align}
where we have dropped the subscripts for $\beta_{e}$ and $\kappa_{e}$ because there is only a single internal edge. The goal is now to compute $f(\eta,\xi)$ in Eq. (\ref{asymptoticlimitx4}). 

We now use the same computation strategy as in Section \ref{warmupgkz} to evaluate the integral. We apply the Mellin-Barnes transformation to both the $\mathcal{G}$ polynomial, but not the $(\beta+\kappa)$-polynomial. The primary difference from the previous section is that not all variables are integrated from zero to infinity. We integrate the $\beta$ variable from $\kappa$ to infinity. To address this issue, we make the u-sub $\beta\rightarrow \kappa \beta$ so $\beta$ is integrated from one to infinity and then use the identity:
\begin{equation}
\int_{1}^{\infty}\frac{d\beta}{\beta} (\beta-1)^{B} \beta^{A}=\frac{\Gamma [B+1] \Gamma [-A-B-1]}{\Gamma [-A]} \ . 
\end{equation}
The result is that the integral over $\beta$ does not introduce a delta-function, but instead adds a beta function in the residue. In principle, we could have made the computation the same as the previous section by making the u-sub $\beta\rightarrow \beta'+\kappa$ so all integration variables are integrated from zero to infinity. Unfortunately, this u-sub introduces a large number of independent terms in $\mathcal{G}$. As emphasized in the previous section, it is very undesirable to have a large number of terms in $\mathcal{G}$ because it means we need to evaluate a greater number of nested sums. The other integrals remain the same. 

The relevant $\mathcal{A}$ matrix is then 
\begin{equation}
\mathcal{A}=\left(
\begin{array}{ccccccccccccccccccccc}
 1 & 1 & 1 & 1 & 1 & 1 & 1 & 1 & 1 & 1 & 1 & 1 & 1 & 1 & 1 & 1 & 1 & 1 & 1 & 1 & 1 \\
 1 & 0 & 1 & 0 & 1 & 1 & 0 & 0 & 1 & 1 & 0 & 0 & 1 & 0 & 0 & 1 & 0 & 0 & 0 & 0 & 0 \\
 0 & 1 & 1 & 0 & 0 & 0 & 1 & 1 & 1 & 0 & 1 & 0 & 0 & 1 & 0 & 0 & 1 & 0 & 0 & 0 & 0 \\
 0 & 0 & 0 & 1 & 1 & 1 & 1 & 1 & 1 & 0 & 0 & 1 & 1 & 1 & 0 & 0 & 0 & 0 & 0 & 0 & 1 \\
 0 & 0 & 0 & 0 & 0 & 0 & 0 & 0 & 0 & 1 & 1 & 1 & 1 & 1 & 0 & 0 & 0 & 0 & 0 & 0 & 0 \\
 1 & 1 & 1 & 1 & 0 & 1 & 0 & 1 & 0 & 1 & 1 & 1 & 0 & 0 & 1 & 0 & 0 & 1 & 1 & 1 & 0 \\
 0 & 0 & 0 & 0 & 0 & 0 & 0 & 0 & 0 & 0 & 0 & 0 & 0 & 0 & 2 & 2 & 2 & 1 & 1 & 0 & 0 \\
 0 & 0 & 0 & 0 & 0 & 0 & 0 & 0 & 0 & 0 & 0 & 0 & 0 & 0 & 0 & 0 & 0 & 1 & 1 & 2 & 2 \\
\end{array}
\right)
\end{equation}
where the $i$th row corresponds to the exponents of the $(i+1)$th variable in the list $(\alpha_{1},\alpha_{2},\alpha_{3},\alpha_{4},\kappa,z_{1},z_{2})$ for all terms in the $\mathcal{G}$ polynomial.\footnote{Note that we have already performed the re-scaling $\beta \rightarrow \beta \kappa$.} The relevant $\beta$-vector is 
\begin{equation}
\beta= \begin{bmatrix}
\frac{1}{2}(d-\Delta^{\textrm{ext}}-2\Delta_{5}) \\
 -\Delta_{1}\\
 -\Delta_{2}\\
 -\Delta_{3}\\
 -\Delta_{4}\\
 -\Delta_{5} \\
 d-\Delta_{1}-\Delta_{2}-\Delta_{5}\\
  d-\Delta_{3}-\Delta_{4}-\Delta_{5}
\end{bmatrix}
\end{equation}
such that the sum over residues corresponds to summing over families of solutions to 
\begin{equation}\label{constraintequ}
\beta=\mathcal{A}\cdot n 
\end{equation}
Each solution corresponds to taking $21-8=13$ sequential residues on the support of Eq. (\ref{constraintequ}). Unfortunately, the number of nested sums makes the practical evaluation of the expression very difficult. Even if we only keep the first $m$ contributions from each sum, it still leads to $m^{13}$ terms. Therefore, while we can algorithmically derive a series expression for any Witten diagram, actually evaluating this series appears computationally intractable for higher-loop Witten diagrams without additional simplifications. 

\section{Discussion}\label{discussion}

In this paper, we introduced a new parameterization of generic Witten diagrams that directly generalizes the Feynman parameterization of flat-space amplitudes. Unlike the flat-space case, the curved-space analog involves three distinct polynomials. In the position-space WF parameterization, we demonstrated how these polynomials arise from specific subgraphs of the Witten diagram. In contrast, in the momentum-space formulation, two of them can be written in terms of familiar flat-space Symanzik polynomials. This identification allowed us to prove a generalization of Weinberg’s theorem on the UV convergence of Feynman integrals~\cite{Weinberg:1959nj}. We then showed how one can, in principle, derive series representations for generic Witten diagrams. These results serve as a proof of concept of the utility of the WF parameterization.  In general, the WF parameterization opens the door to many future applications; any computational technique for flat-space Feynman integrals that relies on Feynman parameterization can now be extended to Witten diagrams using the WF parameterization.

For dS, we focused on computing correlators using the Schwinger-Keldysh in-in formalism~\cite{Schwinger:1960qe,Keldysh:1964ud}. Alternatively, one could compute wavefunction coefficients~\cite{Hartle:1983ai,Harlow:2011ke,Anninos:2014lwa,Maldacena:2002vr}, following a procedure similar to that in Section~\ref{desitterwfpos} but with different propagators. For the wavefunction, the relevant appropriate propagators satisfy Dirichlet boundary condition at some late time~$\eta_{f}$ and Bunch-Davies boundary condition in the far past.\footnote{We thank Nima Arkani-Hamed and Carolina Figueiredo for discussions on this point.} To obtain the propagator for the wavefunction coefficients, one adds a homogeneous term to the dS time-ordered propagator in Eq.~(\ref{timeprop}) that enforces the Dirichlet boundary condition~\cite{Liu:2024str}. It is currently unclear whether this homogeneous term, or its sum with the dS propagator in Eq.~(\ref{timeprop}), admits a Schwinger representation for generic masses, as the time-ordered dS propagator does in Eq.~(\ref{dSGrep}).

For pedagogy, we focused on Witten diagrams of scalars with generic masses and non-derivative couplings. However, the generalization to derivatively coupled scalars is trivial; just act the vertex derivatives on the bulk-to-bulk and bulk-to-boundary propagators. In addition, the generalization to spinning propagators should follow immediately from the formula for spinning propagators in terms of ${}_{2}F_{1}$-functions in Ref. \cite{Costa:2014kfa}, see also Ref. \cite{Sleight:2017fpc}.

Our analysis focused on the UV structure of divergences, and we have not yet examined IR divergences. Both the UV and IR structures of Feynman integrals are manifest in standard Feynman parameterizations \cite{Hepp:1966eg,Speer:1975dc,Arkani-Hamed:2022cqe}. The WF parameterization could be helpful for computations in dS space involving light, non-derivatively coupled scalars, which are known to exhibit IR divergences and resummation \cite{Burgess:2009bs,Burgess:2010dd,Gorbenko:2019rza}. 

While we outlined a basic procedure for converting the WF representation into a series form, we did not provide a tractable series expansion for any previously unsolved Witten diagram. Moreover, although our examples suggest that a convergent series representation exists for any given kinematic point, we did not establish this more generally, even for the s-channel diagram. More broadly, we expect that a variety of computational techniques could be developed to reduce the number of nested summations required for a given diagram. The WF parameterized integrals are not generic generalized Euler integrals, but instead exhibit additional structure that should be leveraged. We expect the methods introduced in Refs.~\cite{Arkani-Hamed:2024nzc,Xianyu:2023ytd,Raman:2025tsg} are likely to be especially useful in this direction. 

Several alternative methods have been developed to study Witten diagrams, including Mellin-space techniques~\cite{Mack:2009mi,Mack:2009gy,Penedones:2010ue,Fitzpatrick:2011ia,Sleight:2019mgd} and representations of correlators as differential operators acting on contact diagrams~\cite{Roehrig:2020kck,Eberhardt:2020ewh,Diwakar:2021juk,Herderschee:2021jbi}. These approaches have yielded valuable insights into the structure of perturbation theory in curved spacetimes. However, translating such formal results into explicit expressions for position- or momentum-space correlators, especially those of potential experimental relevance, often remains highly non-trivial~\cite{Diwakar:2021juk,Herderschee:2022ntr}. Nonetheless, it would be interesting to clarify the relationship between the WF parameterization and these alternative frameworks. 

Tropical Monte Carlo offers a promising approach to the numerical evaluation of Witten diagrams. Feynman parameterized integrals are well suited to such methods \cite{Borinsky:2020rqs}, and the WF parameterization should likewise be amenable to tropical techniques. In momentum space, the $\mathcal{U}$ and $\mathcal{F}$ polynomials can be evaluated efficiently using their expression as determinants of reduced Laplacian matrices, enabling fast numerical algorithms \cite{spielman2014nearly}. In position space, the determinant representations of the $\mathcal{W}$ and $\mathcal{Y}$ polynomials in Eq.~(\ref{deterform}) similarly allow for efficient computation. Once a Cholesky decomposition of the associated $V \times V$ matrix is obtained, which scales as $\mathcal{O}(V^3)$, the determinant can be evaluated in linear time.\footnote{We thank Michael Borinsky for discussion on this point.} However, even this method is likely suboptimal, and we expect more efficient algorithms may exist that exploit the structure of the matrix. The $\mathcal{Z}$ polynomial, by contrast, cannot be expressed as a determinant, but its evaluation remains inexpensive, as the number of terms in its sum scales linearly with the number of edges. Beyond these considerations, we find in all examples that the Newton polytopes of the $\mathcal{Y}$ and $\mathcal{W}$ polynomials are generalized permutahedra \cite{postnikov2009permutohedra}, a structure that should prove especially useful for tropical Monte Carlo. Unfortunately, the $\mathcal{Z}$ polynomial does not share this feature.

A very important computational tool for Feynman integrals is differential equations. Given a first order differential equation for the integral, one can use numerical analysis to derive the value of integral anywhere in parameter space given an accurate evaluation at some initial point. The differential and difference equations obeyed by Feynman integrals can be derived using the Feynman parameterized integral \cite{Weinzierl:2022eaz,Matsubara-Heo:2023ylc}. We expect similar computations are possible for Witten diagrams. We plan to leverage D-module theory \cite{saito2013grobner,sattelberger2019d}  in the future to algorithmically compute the Pfaffian systems of higher loop Witten diagrams. We expect modern computational algebraic geometry packages, in particular Ref. \cite{ConnectionMatricesSource}, to be very useful. 

Finally, the method of regions \cite{Smirnov:1991jn,Beneke:1997zp,Smirnov:1998vk,Smirnov:1999bza,Pak:2010pt,Ananthanarayan:2018tog,Ananthanarayan:2020ptw} provides a very general algorithm for evaluating generalized Euler integrals when external parameters are taken parametrically large or small. We leveraged the method of regions to simplify the conformal Witten diagrams in Section \ref{levconf}. However, we could also consider studying different physical kinematic limits, such as the OPE limit in Ref. \cite{Arkani-Hamed:2015bza}. For dS correlators, one idea would be to search for limits where the signal-to-noise ratio is enhanced, improving the measurement prospects. Alternatively, it could be interesting to study flat space limits \cite{Paulos:2016fap,Dubovsky:2017cnj,Hijano:2019qmi,Li:2021snj}, which are known to exhibit some subtleties \cite{Komatsu:2020sag}.\footnote{We thank Yue-Zhou Li for discussion on this point.}


\subsection*{Acknowledgments}

We thank Nima Arkani-Hamed, Michael Borinsky, Carolina Figueiredo, Yue-Zhou Li, Qianshu Lu, Radu Roiban, and Giulio Salvatori for helpful discussions. We thank Aaron Hillman and Beatrix Muehlmann for constructive comments on the draft. AH is grateful to the Simons Foundation as well as the Edward and Kiyomi Baird Founders’ Circle Member Recognition for their support.

\appendix

\section{Deriving the Schwinger parameterization of the 2F1 function}\label{dersch2f1}

To derive the Schwinger parameterization (\ref{scwhingerpre}) from the propagator in Eq. (\ref{adesitterprop}), we derive a related formula for the ${}_{2}F_{1}$ hypergeometric function.

We start with the well known formula for the ${}_{2}F_{1}$ hypergeometric function:
\begin{equation}
{}_{2}F_{1}[a,b,c;z]=\frac{\Gamma[c]}{\Gamma[b]\Gamma[c-b]}\int_{0}^{1} dt t^{b-1}(1-t)^{c-b-1}(1-t z)^{-a} \ .
\end{equation}
We insert the delta-function 
\begin{equation}
0<t<1:\quad 1=\int_{0}^{\infty} dr \delta(1-t-r)
\end{equation}
and re-arrange the integrand on the support of the delta function
\begin{equation}
{}_{2}F_{1}[a,b,c;z]=\frac{\Gamma[c]}{\Gamma[b]\Gamma[c-b]}\int_{0}^{\infty} dt dr \delta(1-r-t) t^{b-1}r^{c-b-1}(r+t(1-z))^{-a} 
\end{equation}
where we have extended the integration region for $t$ from $t\in (0,1)$ to $t\in (0,\infty)$. We then re-write the component containing $z$ into an exponential 
\begin{equation}
{}_{2}F_{1}[a,b,c;z]=\frac{\Gamma[c]}{\Gamma[b]\Gamma[c-b]\Gamma[a]}\int_{0}^{\infty} d\lambda \lambda^{a-1} \int_{0}^{\infty} dt dr \delta(1-r-t) t^{b-1}r^{c-b-1}e^{-\lambda (r+t(1-z))}
\end{equation}
and now re-scale $r\rightarrow r/\lambda$ and $t\rightarrow t/\lambda$, which yields 
\begin{equation}
\begin{split}
{}_{2}F_{1}[a,b,c;z]=\frac{\Gamma[c]}{\Gamma[b]\Gamma[c-b]\Gamma[a]}\int_{0}^{\infty} &d\lambda \lambda^{a-c} \int_{0}^{1} dt dr \delta(\lambda-r-t) \\
&\times t^{b-1}r^{c-b-1}(r+t)^{A}e^{-(r+t(1-z))}
\end{split}
\end{equation}
We can integrate out $\lambda$, leaving the final result 
\begin{equation}\label{formula2f1}
{}_{2}F_{1}[a,b,c;z]=\frac{\Gamma[c]}{\Gamma[b]\Gamma[c-b]\Gamma[a]} \int_{0}^{\infty} dt dr t^{b-1}r^{c-b-1}(r+t)^{a-c}e^{-(r+t(1-z))} \ .
\end{equation}
Using the above formula for the ${}_{2}F_{1}$ functions that appear in the two-point propagator (\ref{adesitterprop}) and performing the u-sub
\begin{equation}
t\rightarrow ((z-z')^{2}+(x-x')^{2}) \kappa, \quad r\rightarrow ((z-z')^{2}+(x-x')^{2}) (\beta-\kappa)
\end{equation}
leads to Eq. (\ref{scwhingerpre}).

\bibliographystyle{apsrev4-1long}
\bibliography{GeneralBibliography.bib}
\end{document}